\documentclass[aps,pre,twocolumn,superscriptaddress,amsmath,amssymb,floatfix]{revtex4-1}

\usepackage{graphicx}
\usepackage[usenames]{color} 
\usepackage{comment} 
\usepackage{bm}
\usepackage{amsmath}

\newcommand{\be}{\begin{equation}} 
\newcommand{\ee}{\end{equation}}
\newcommand{\bea}{\begin{eqnarray}} 
\newcommand{\eea}{\end{eqnarray}}
\newcommand{\br}{{\bf r}}
\newcommand{\rhoh}{\hat{\rho}}

\begin{document}

\title{Drying and Wetting Transitions of a Lennard-Jones Fluid: Simulations and Density Functional Theory}

\author{Robert Evans}  
\affiliation{H. H. Wills Physics Laboratory, University of Bristol, Royal Fort, Bristol BS8 1TL, United Kingdom}
\author{Maria C. Stewart}  
\affiliation{H. H. Wills Physics Laboratory, University of Bristol, Royal Fort, Bristol BS8 1TL, United Kingdom}
\author{Nigel B. Wilding} 
\affiliation{Department of Physics, University of Bath, Bath BA2 7AY,
United Kingdom} 

\begin{abstract}

We report a theoretical and simulation study of the drying
and wetting phase transitions of a truncated Lennard-Jones fluid at a flat
structureless wall. Binding potential calculations predict that the
nature of these transitions depends on whether the wall-fluid
attraction has a long ranged (LR) power law decay, or is instead truncated,
rendering it short ranged (SR). Using grand canonical Monte
Carlo simulation and classical density functional theory we examine
both cases in detail.  We find that for the LR case wetting is first
order, while drying is continuous (critical) and occurs exactly at
zero attractive wall strength, ie. in the limit of a hard wall.  In the SR case,
drying is also critical but the order of the wetting transition
depends on the truncation range of the wall-fluid potential.  We
characterize the approach to critical drying and wetting in terms of
the density and local compressibility profiles and via the finite-size
scaling properties of the probability distribution of the overall
density.  For the LR case, where the drying point is known exactly,
this analysis allows us to estimate the exponent $\nu_\parallel$ which
controls the parallel correlation length, i.e. the extent of vapor
bubbles at the wall.  Surprisingly, the value we obtain  is over twice that
predicted by mean field and renormalization group calculations,
despite the fact that our three dimensional system is at the upper
critical dimension where mean field theory for critical exponents is expected to
hold. We suggest reasons for this discrepancy.

\end{abstract}

\maketitle

\section{Introduction}

\begin{figure}[h]
\includegraphics[type=pdf,ext=.pdf,read=.pdf,width=0.94\columnwidth,clip=true]{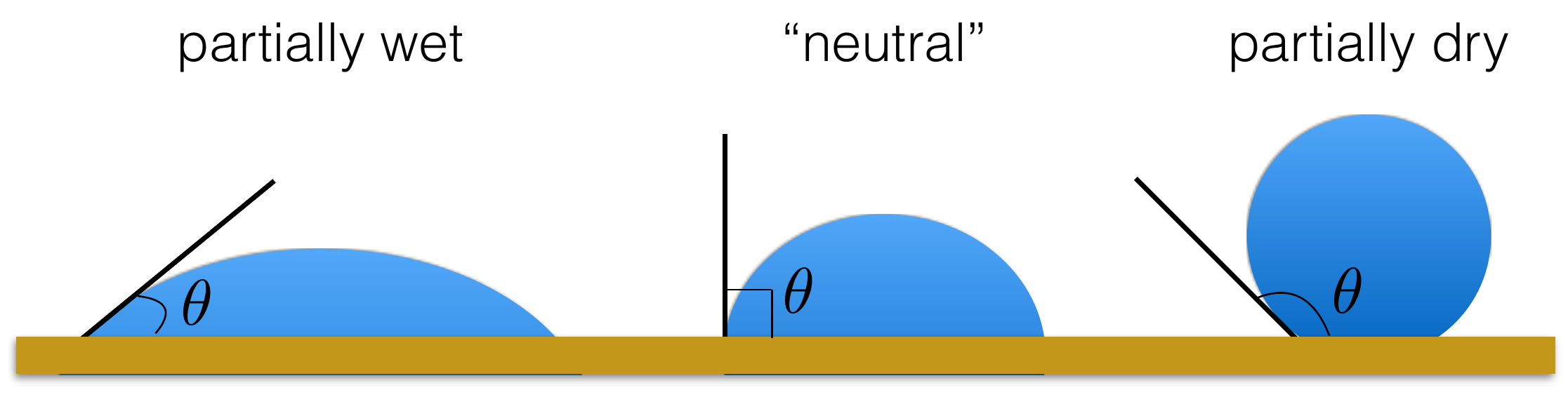}

\caption{Schematic diagram of a liquid drop on a flat surface showing how the contact angle $\theta$ 
varies as the attractive strength of the wall-fluid interaction potential is reduced from a large value (left) to a small value (right). The `neutral' case $\theta=90^\circ$ marks the boundary between the partially wet and partially dry regimes. }

\label{fig:contact_angle_schem}
\end{figure}

The behaviour of a liquid drop in equilibrium with its vapor and in contact with a flat substrate (or `wall') is characterised in thermodynamic terms by
the contact angle $\theta$ that the drop makes with the substrate\,\cite{deGennes:1985aa}. The precise value of $\theta$
depends on the surface chemistry of the substrate, but in broad terms, strong wall-fluid attraction is associated with a
small contact angle, while weak attraction is associated with a large contact angle, as shown schematically in
Fig.~\ref{fig:contact_angle_schem}. On increasing the wall-fluid attraction the contact angle approaches the limit
$\theta \to 0^\circ$. This corresponds to the wetting transition in which a
macroscopic liquid layer intrudes between the wall and the vapor. The drying transition is the counterpart of wetting
that occurs as the wall attraction is progressively weakened so that $\theta\to 180^\circ$, whereupon a macroscopic vapor
layer intrudes between the wall and the liquid. Values of $0^\circ <\theta<90^\circ$ are often termed partially wet, while
values $90^\circ <\theta<180^\circ$ are termed partially dry. Young's equation: 
 
\be
\gamma_{vl}\cos(\theta) = \gamma_{wv}-\gamma_{wl}\:,
\label{eq:Young}
\ee 
expresses $\theta$ in terms of the wall-vapor ($wv$), wall-liquid
($wl$) and vapor-liquid ($vl$) surface tensions.

Let us consider first the nature of wetting. This has been the subject of enduring experimental, theoretical and
simulation interest~\cite{Bonn:2009if}. As is well established, the character of the transition can be either
discontinuous (first order) or continuous (critical) depending in a subtle fashion on whether the wall-fluid ($wf$)
potential is long-ranged (LR) or short ranged (SR) and on whether the fluid-fluid ($ff$) interaction is LR or SR.
\cite{Nightingale:1983ty,Nightingale:1984aa,Dietrich:1985rz,Ebner:1985aa,Ebner:1987aa,Ebner:1987xy,Binder:1986aa,Ross:2001aa}. Most
experimental studies of wetting transitions find these to be first order (see \cite{Bonn:2001aa} for a review and
\cite{Friedman:2013nx} for a recent study of water). Critical wetting is much rarer (although potentially more
interesting from a fundamental perspective) and experimental reports have, to date, been limited to a few special cases
\cite{Ragil:1996aa,Ross:1999aa,Ross:2001aa}. Consequently, the bulk of progress towards elucidating the character of
critical wetting has come from theoretical and simulation approaches, which have revealed a wealth of complex
behaviour\,
\cite{Nakanishi:1982aa,Brezin:1983dn,Nightingale:1983ty,Nightingale:1984aa,Dietrich:1985rz,Ebner:1985aa,Ebner:1987xy,Binder:1986aa,Binder:1989aa,Ragil:1996aa,Ross:1999aa,Evans:1990aa,Evans:1989aa,Parry:1991zm,Parry:2008rp,Parry:2008ef,Albano:2012db,Bryk:2013pi}.
Theoretical treatments typically adopt a mean field (MF) approach based on Landau theory, binding potentials or density
functional theory (DFT). These predict that critical wetting occurs when both $wf$ and $ff$ interactions are SR, and
that $d=3$ is the upper critical dimension. Hence three dimensional systems are a borderline case for the validity of
mean field theory. Renormalization Group (RG) calculations \cite{Brezin:1983dn,Fisher:1985aa} predict non-universal
behaviour in this instance-- a finding that has prompted concerted simulation efforts to clarify the nature of the
criticality. Unfortunately simulations are complicated by the finite-size effects that stem from the divergent critical
correlations. Early work for a nearest neighbor Ising model with surface fields \cite{Binder:1986aa,Binder:1989aa}
yielded pronounced discrepancies with RG predictions regarding the critical exponents, prompting efforts to extend the
theoretical framework to non-local interfacial Hamiltonians \cite{Parry:2008ef,Parry:2008rp,Parry:2009aa} in order to
explain the differences. More recent simulation studies have attempted to deploy finite-size scaling (FSS) techniques,
using data from a wide range of lattice sizes, in order to pinpoint the transition and elucidate its character
\cite{Albano:2012db,Bryk:2013pi}. Significant inaccuracies were identified in the earlier estimates of the transition
point, which had in turn skewed the estimates of critical point properties. However, difficulties in applying the FSS
methodology to three dimensional systems were also reported \cite{Bryk:2013pi} and thus the precise relationship between
simulation data and theoretical predictions is arguably not settled fully.

As indicated above, simulation studies of critical wetting have been confined exclusively to Ising models on account of
their computational tractability. However, it is interesting to ask to what extent realistic fluid models share the
properties of the lattice systems. Although fluid-magnet universality suggests that the critical scaling properties
should be identical, the particle-hole symmetry inherent in Ising models is expected to engender features that do not
occur in real fluids. For example wetting is formally equivalent to drying for SR surface magnetic fields in the Ising
model, but there is no reason to think that wetting and drying are equivalent for realistic fluid models. Furthermore
it is unclear how to translate some of the model parameters commonly employed in the Ising context, such as enhanced
surface layer couplings and surface magnetic fields, to the case of realistic fluids in which the substrate-fluid
interactions are fully described by the associated wall potential. What is clear, however, is that simulations of
realistic fluids can be expected to be considerably more computationally demanding than for Ising models, and
accordingly one should expect that the range of accessible system sizes is correspondingly smaller.

In common with wetting, relatively little is known about the fundamental nature of drying transitions in realistic fluids.
Experimental studies are far scarcer than for wetting owing to the challenges of fabricating
substrates for which the contact angle of a liquid drop is large. For instance, for water on `hydrophobic' surfaces such as
Teflon, wax or self-assembled monolayers, the contact angle does not generally exceed $\theta \approx
130^\circ$. While this precludes detailed study of the approach to the drying transition, interesting effects have
nevertheless been reported for strongly hydrophobic substrates, such as a depletion region of one or two molecular layers adjacent to the substrate in which the one-body density
is considerably reduced compared to its bulk value \cite{Mezger:2006zl, Ocko:2008fv,
Mezger:2010lq,Chattopadhyay:2010aa,Chattopadhyay:2011aa,Uysal:2013aa}. Encouragingly, the ability to study the
drying transition in detail is likely to improve in the future with the advent of novel nano and
micro-structured surfaces which exhibit contact angles approaching $170^\circ$
\cite{Lafuma:2003aa,Quere:2005cl,Simpson:2015le,Li:2007aa,Ueda:2013aa}. Such `superhydrophobic' surfaces are
of widespread interest for their potential technological applications, including self cleaning surfaces and
chemical separation processes \cite{Simpson:2015le}.

On the theoretical side, it is well established that complete drying occurs for any liquid (that exhibits liquid-vapor
coexistence) adsorbed at a planar hard wall, see e.g. \cite{Henderson:1985aa,Oettel:2005aa}. However, the situation for
attractive $wf$ interactions is less clear. Early work emphasized the central role of the range of the potentials
\cite{Ebner:1987aa,Ebner:1987xy} and reported that for a SR lattice-gas system with LR $wf$ interactions, no drying
transition can occur, while any wetting transition is first order. This important prediction is specific to the
lattice-gas (Ising) model. We reexamine this in the context of a fluid in contact with a hard wall plus attractive LR
tail.

On the simulation front, a number of studies have been performed, but no clear consensus regarding the nature of
drying has yet emerged. Monte Carlo simulation studies of a Lennard-Jones liquid \cite{Oleinikova:2005uo,Brovchenko:2005ph,Brovchenko:2004ek} utilizing LR $wf$ interactions
 reported no signs of a drying transition, stating this was in
accord with the theoretical predictions\,\cite{Ebner:1987xy}. Molecular Dynamics (MD) studies by two separate groups, of
both truncated Lennard-Jones (LJ) and square-well fluids reported a drying transition at
some small but non-zero strength of SR $wf$ attraction, but disagreed regarding its character, with one group claiming that
the transition is first order \cite{Henderson:1990nq,Henderson:1992kk,Swol:1989by,Swol:1991fq} and the other
\cite{Nijmeijer:1992fk,Nijmeijer:1991sw,Bruin:1995ud,Bruin:1998aa} that it is critical. (Note that DFT calculations came out in favor of critical drying \cite{Henderson:1992kk}.) More recently Monte Carlo (MC)
studies of the contact angle in a LJ fluid  \cite{Rane:2011ly} and SPC/E water \cite{Kumar:2013aa} at various types of substrate
pointed tentatively to a critical drying transition for the LR $wf$ case. Hints of differences in
the character of the approach to drying between systems with LR and SR wall-fluid interactions were also noted. Separate
MD studies of a model for water at a weakly attractive substrate found evidence for enhanced density fluctuations in the
surface region\,\cite{Chandler:2007aa,Patel:2010dz,Patel:2012aa, Willard:2014aa,Mittal:2010aa,Godawat:2011aa}. The
latter finding was subsequently rationalized by MC simulations for SPC/E water which provided firm evidence that the drying
transition in water is critical and that this fact is responsible for the enhanced fluctuations in the surface
region\,\cite{Evans:2015wo}.

Despite substantial progress made in understanding the physics of wetting and drying in realistic fluids, a number of
fundamental and practical questions remain unanswered. Principal among these are: i) What characteristics of the model
system are responsible for determining the order of wetting and drying transitions? ii) How can wetting and drying
points be located accurately via computer simulation? iii) If a transition is critical, what is the nature of the
near-critical fluctuations and the values of the critical exponents and how can these be measured accurately? iv) What
is the role of finite-size effects in characterizing surface phase transitions? 

In the present contribution we address
these issues using a combination of theoretical and computational techniques. For simulational expediency we consider $ff$ interactions that are exclusively SR
in nature (a truncated LJ potential),  but for the $wf$ interactions we consider both the LR and SR cases. Our focus is mainly on the drying transition in systems with LR $wf$ interactions because:
(a) drying turns out to be critical for LR $wf$; (b) the drying point occurs at exactly zero attractive
wall strength (ie in the limit of a hard wall) -- a feature which allows us to study the transition free from uncertainty
regarding its location. However, we also report results for drying in the case when the $wf$ interactions are SR and for
wetting in the case of LR and SR $wf$ interactions. The latter case of wetting for SR $wf$ is also found to be critical but the theoretical
predictions for the nature of the criticality are different from those for drying with LR $wf$ interactions.

The layout of the paper is as follows\,\footnote{A short account of some of our results has previously been published
elsewhere \cite{Evans:2016aa}}. Section~\ref{sec:modelmeth} describes our model of a Lennard-Jones fluid confined
between smooth parallel walls. In Section~\ref{sec:bg} we set out some general features of adsorption, surface phase
behavior and criticality for a simple fluid. Section.~\ref{sec:theory} provides details of the theoretical methods we have used to study the model, namely MF and RG analysis of a binding potential and classical DFT. The grand canonical Monte Carlo (GCMC) simulation methods are described in Section~\ref{sec:simmeth}. Our results for the drying and wetting
properties are set out in Sections~\ref{sec:dftresults} and \ref{sec:simresults}, and are discussed in
Section~\ref{sec:discuss}.

\section{Model Fluids}
\label{sec:modelmeth}

Our grand canonical Monte Carlo (GCMC) simulations consider the drying and wetting behaviour of a Lennard-Jones (LJ) fluid in which particles interact via the potential,
\be
\phi_{\rm att}(r)=\left \{ \begin{array}{ll}
 4\epsilon_{LJ}\left[\left(\frac{\sigma}{r}\right)^{12}-\left(\frac{\sigma}{r}\right)^{6}\right], & r\le r_c \, ,\\
0, & r>r_c,\\
\end{array}
\right.
\label{eq:Simpot}
\ee
with $\epsilon_{LJ}$ the well-depth of the potential and $\sigma$ the
LJ diameter. We choose $r_c=2.5\sigma$, for which criticality occurs
\cite{Wilding1995} at $k_BT_c=1.1876(3) \epsilon_{LJ}$. We work at
$k_BT=0.91954\epsilon_{LJ}=0.775T_c$ for which coexistence occurs at
$\beta\mu_{co}=-3.865950(20)$, with coexistence densities $\rho_l\sigma^3=0.704(1)$
and $\rho_v\sigma^3=0.0286(2)$; and also at $k_BT=1.0\epsilon_{LJ}=0.842T_c$ for
which $\beta\mu_{co}=-3.457131(25)$, $\rho_l\sigma^3=0.653(1),\rho_v\sigma^3=0.0504(3)$, $\beta=(k_BT)^{-1}$.
The choice of cutoff $r_c$ is motivated computationally and follows that of most of the LJ
community. 

The fluid is confined within a slit pore comprising 
 two planar walls of area $(L\sigma)^2$ separated by a distance $D\sigma$, so that the volume is
 \be
 V=(L\sigma)^2D\sigma\:.
 \label{eq:slitvol}
 \ee
 Periodic boundary conditions are applied in the directions parallel to the planar walls.
We employ three types of wall-fluid potential in our GCMC
simulations. The SR potential for a single wall is a square-well given by

\be
W_{\rm SR}(z)=\left \{ \begin{array}{ll}
\infty, \mbox{\hspace{4mm}}   &  z\le 0  \\
 -\epsilon, & 0<z<\sigma/2 \, ,\\
0, & z>\sigma/2,\\
\end{array}
\right.
\label{eq:SRpot}
\ee
where $\epsilon$ is the well-depth. 

Two types of LR potential are considered in this work. The first is given by the
well known $9$-$3$ form having a decaying attractive part at large $z$
and a steep repulsive part at small $z$:

\be
W_{\rm LR}(z)=\left \{ \begin{array}{ll}
\infty, \mbox{\hspace{4mm}}   &  z\le 0  \\
\epsilon_w\epsilon_{LJ}\left[\frac{2}{15}\left(\frac{\sigma}{z}\right)^9-\left(\frac{\sigma}{z}\right)^3\right], & z>0 \, ,\\\end{array}
\right.
\label{eq:LRpot}
\ee
where $\epsilon_w$ is a dimensionless measure of the strength of the wall-fluid
attraction. At the minimum of (\ref{eq:LRpot}) the value of the wall-fluid potential is
$-1.0541\epsilon_w\epsilon_{LJ}=-\epsilon$.

The second LR potential, a modification of the first, is obtained by making the replacement $z\to{\tilde z}$ in
(\ref{eq:LRpot}) with ${\tilde z} = z+(2/5)^{1/6}\sigma$. This shifts the minimum of the $9$-$3$ potential to the hard
wall at $z=0$, leading to an infinitely steep repulsive part. The motivation for utilizing the modified form is that
when studying a slit geometry (see below) the wall separation and hence the slit volume is unambiguously defined for all
$\epsilon_w$. This is not the case for the standard form (\ref{eq:LRpot}) in the regime of interest for drying, namely
$\epsilon_w\approx 0$. As $\epsilon_w\to 0$ the effective wall position (given by the value of $z$ for which the repulsive
energy is $\sim k_BT$) shifts strongly, leading to artifacts in measurements of the total number density. The GCMC
results reported in Sec.~\ref{sec:simresults} are for the modified potential. The DFT results in
Sec.~\ref{sec:dftresults} are for (\ref{eq:LRpot}).

\section{Background to Wetting, Drying and Confinement with Planar Walls}
\label{sec:bg}

\subsection{Statistical Mechanics of Adsorption: Thermodynamics and Correlation Functions}
\label{sec:adsoprtion}
In this subsection we summarize key results in the statistical mechanics of adsorption pertinent to our GCMC and DFT investigations. For convenience we consider the fluid to be adsorbed at a single (planar) wall of (infinite) interfacial area $A$ exerting a potential of the type (\ref{eq:SRpot},\ref{eq:LRpot}). Then the average one-body density $\rho(\br) = \rho(z) =0, z<0$. Extension to the fluid confined by two walls is straightforward. As is appropriate for adsorption studies, we work grand canonically with a reservoir at fixed chemical potential $\mu$ and temperature $T$. Thus

\be
\rho(\br)=\langle \rhoh (\br)\rangle \equiv\langle \sum_{i=1}^N\delta (\br-\br_i)\rangle
\ee
where the brackets $\langle\rangle$ denote a GC average and we introduced the usual particle density operator for $N$ particles with coordinates ${\bf r}_i$. Two-body correlations are described by the density-density correlation function:
\bea
G ( \br_1 , \br_2 )&\equiv&\langle (\rhoh(\br_1)-\langle\rhoh(\br_1)\rangle)(\rhoh(\br_2)-\langle\rhoh(\br_2)\rangle)\rangle\nonumber\\
&=&G(z_1,z_2;R)
\eea
where $R= \sqrt{(x_1 -x_2 )^2 +(y_1 -y_2 )^2}$ is the transverse separation between particles (atoms). Connection with surface thermodynamics is made via the Gibbs adsorption equation

\be
\Gamma \equiv  \int_0^\infty dz (\rho(z) -\rho_b ) = -\frac{1}{A}\left(\frac{\partial \Omega_{ex}}{\partial\mu}\right)_T
\label{eq:Gammadef}
\ee
i.e. $\Gamma$, the excess number of particles per unit area, is the minus of the derivative of the
excess grand potential $\Omega_{ex}$ w.r.t.~$\mu$. $\rho_b \equiv \rho_b (\mu,T )$ is the density of the bulk fluid far from the wall and the excess quantity is defined by $\Omega_{ex} \equiv \Omega + pV$ where $\Omega$ is the total grand potential, $V$ is the accessible volume and $p \equiv p(\mu,T )$ is the pressure of the bulk 
fluid.

The second derivative of $\Omega_{ex}$ w.r.t.~$\mu$ yields the surface excess compressibility, and this can be written as a fluctuation formula \cite{Evans:1987aa,Evans:2015aa}
\be
\chi_{ex} \equiv \left(\frac{\partial\Gamma}{\partial \mu}\right)_T=\frac{\beta}{A}[(\langle N^2\rangle-\langle N\rangle^2)-(\langle N_b^2 \rangle-\langle N_b \rangle^2)] 
\label{eq:chiex1}
\ee
where it is implied that the surface area $A \to\infty$. The first term in (\ref{eq:chiex1}) is the mean-square fluctuation in the total number of particles, which must be positive to ensure stability. The second is the corresponding quantity for the bulk fluid at the same $(\mu, T)$. The difference can be negative \cite{Evans:1987aa}. Bratko {\em et al} \cite{Bratko:2007aa,Bratko:2010aa}  measured the first term of (\ref{eq:chiex1}) and, more recently, Kumar and Errington \cite{Kumar:2013kx} measured $\chi_{ex}$  in GCMC simulations of SPC/E water at hydrophobic substrates. Whilst $\chi_{ex}$ provides a measure of the overall compressibility of the adsorbed fluid and the strength of fluctuations in the total number of particles, this quantity does not provide information about the spatial location of the important density fluctuations, i.e. at which distances $z$ from the wall these are most pronounced. Below we define the local compressibility $\chi(z)$ which does provide the appropriate measure.

First we introduce a further sum rule relating a thermodynamic quantity to an (integrated) microscopic quantity. Following \cite{Henderson:1986kx, Evans:1989aa} we suppose that $W (z)$ is such that $\partial W (z) / \partial\epsilon \equiv W_\epsilon (z) / \epsilon$ is independent of the well-depth $\epsilon$. (The wall-fluid potentials we consider here meet this requirement.) The parameter $\epsilon$ acts as a thermodynamic field with a conjugate density $\Theta$. Surface thermodynamics follows from

\be
\frac{1}{A}d(\Omega_{ex})=-sdT-\Gamma d\mu-\Xi d\epsilon\:,
\label{eq:dOmega}
\ee
where $s$ is the surface excess entropy per unit area and the conjugate density is \cite{Evans:1989aa}

\be
\Theta=-\frac{1}{A}\left(\frac{\partial \Omega_{ex}}{\partial\epsilon}\right)_{\mu,T}=-\int_0^\infty dz \rho(z)\frac{W_\epsilon(z)}{\epsilon}
\label{eq:Theta}
\ee

Using the Maxwell relation resulting from (\ref{eq:dOmega}) we obtain the sum rule:

\be
\Gamma_1\equiv\left(\frac{\partial\Gamma}{\partial\epsilon}\right)_T=\left(\frac{\partial\Theta}{\partial\mu}\right)_T\equiv\chi_1
\label{eq:maxwell}
\ee
Eqs. (\ref{eq:Gammadef},\ref{eq:maxwell}) are satisfied identically within the DFT approximations that we employ. We determine $\Gamma$ and $\Theta$ in our GCMC simulations and shall use (\ref{eq:maxwell}) to explore the statistical accuracy of these. Note that the equivalent of (\ref{eq:maxwell}) is well-known in surface (Ising) magnetism where $\Gamma$ is equivalent to the excess magnetization $m_s$, $\epsilon$ plays the role of the locally applied surface magnetic field and $\mu$ plays the role of the bulk magnetic field $h$. For the case of a surface field $h_1$ acting in only the first (surface) layer of spins one has the standard result:

\be
\left(\frac{\partial m_s}{\partial h_1}\right)_T=\left(\frac{\partial m_1}{\partial h}\right)_T
\ee
where $m_1$ is the magnetization in the surface layer. Clearly this particular magnetic case corresponds to a model fluid in which $W \epsilon ) / \epsilon \sim\delta (z)$ so that $\Theta$ reduces to $m_1$. In
the magnetism literature the quantity $(\partial m_1 / \partial h)_T$ is usually termed $\chi_1$, the surface layer susceptibility.

For fluids it is appropriate to define a local compressibility:
\be
\chi(z)\equiv\left(\frac{\partial \rho(z)}{\partial\mu}\right)_T
\label{eq:chizdef}
\ee

This quantity was introduced in early studies of wetting transitions e.g. \cite{Tarazona:1982aa, Evans:1990aa} and shown recently \cite{Evans:2015aa, Evans:2015wo,Evans:2016aa} to provide a valuable measure of the degree of solvophobicity or hydrophobicity of a substrate, i.e. as the well-depth $\epsilon$ decreases and the macroscopic contact angle increases, there is an accompanying increase in the maximum of $\chi(z)$ located at distances $z$ within $1$-$3$ diameters of the substrate. The connection between $\chi(z)$ and density correlations at the substrate is best made by introducing the local, or transverse, structure factor:

\be
S(z_1;q)\equiv\int_{-\infty}^\infty dz_2\int d{\bf R}e^{i{\bf q}\cdot {\bf R}}G(z_1,z_2;R) 
\label{eq:Sq}
\ee
where $q$ is the transverse wave number. $S(z;q)$ provides a measure of the strength and range of transverse correlations at distance $z$ from the substrate and plays an important role in the theory of wetting \cite{Tarazona:1982aa, Evans:1990aa} and in characterizing the structure of the liquid-vapor interface \cite{Parry:2016aa}. It is straightforward to show that $\chi(z)$ is proportional to the $q=0$ limit of (\ref{eq:Sq}):

\be
\chi(z_1)=\beta S(z_1;0)=\beta \int_{-\infty}^\infty dz_2\int d{\bf R}G(z_1,z_2;R) 
\label{eq:chicorr}
\ee
i.e. the local compressibility $\chi(z_1)$ is the integral of the density-density correlation function over the transverse coordinate $R$ and over one normal coordinate, $z_2$. Generally, $\chi(\br)$ can be expressed as a fluctuation formula that follows by differentiating the grand partition function w.r.t. an external potential, to obtain the average one-body density, and then differentiating w.r.t. the chemical potential. One finds \cite{Evans:2015wo}:

\be
\chi(\br)=\beta^{-1} \langle N\rhoh(\br)-\langle N\rangle\langle \rhoh(\br)\rangle\rangle 
\ee
Clearly $\chi(\br)$ is the correlator of the local number density at $\br$ and the total number $N$ of particles.

Finally if we integrate (\ref{eq:chicorr}) over $z_1$ we obtain the surface compressibility sum rule,e.g.~\cite{Nicholson:1982aa, Evans:1989aa}, relating the surface excess compressibility in (\ref{eq:chiex1}) to an integral of the surface structure factor, at $q=0$:

\be
\chi_{ex} =\int_0^\infty dz[\chi (z)-\chi_b]=\int_0^\infty dz[\beta S(z;0)-\chi_b] 
\label{eq:chiex}
\ee
where the bulk contribution $\chi_b = \rho_b^2\kappa_T$; $\kappa_T$ is the usual isothermal compressibility,
proportional to the bulk structure factor at zero wave number.

\subsection{Phenomenology of Wetting, Drying and Surface Criticality}
\label{sec:phenomenology}
Here we remind readers of some of the phenomenology of wetting and drying transitions and describe the critical
exponents that characterize such transitions. Wetting and drying are phase transitions that occur strictly in the limit
of infinite wall area, $A\to\infty$, and for a single wall, i.e infinite wall separation, $D\to\infty$. Using the
adsorption language of the previous subsection, a wetting transition occurs at fixed $T$ when the excess adsorption
$\Gamma$ changes from a finite value to an infinite value as the wall-fluid attraction $\epsilon_w$ is increased to the
transition value $\epsilon_{ww}$; the bulk fluid is a vapor, at fixed chemical potential $\mu=\mu_{co}^-(T)$. For
$\epsilon_w<\epsilon_{ww}$ the (planar) surface tensions satisfy $\gamma_{wv}<\gamma_{wl}+\gamma_{lv}$, where $w$ refers
to wall, $l$ to liquid and $v$ to vapor. From Young's equation (\ref{eq:Young}) it then
follows that $\cos(\theta)<1$ and the situation corresponds to partial wetting in Fig.~\ref{fig:contact_angle_schem}.

For $\epsilon_w>\epsilon_{ww}, \Gamma=\infty, \gamma_{wv}=\gamma_{wl}+\gamma_{lv}$ and $\cos(\theta)=1$, consistent with
the wall-vapor interface being wet by a macroscopically thick layer of liquid. If $\Gamma$ jumps abruptly at the
transition value the transition is first order. If $\Gamma$ diverges continuously at $\epsilon_{ww}$ the transition is
termed continuous or critical. As mentioned in the Introduction, drying is the counterpart of wetting when the bulk
fluid is a liquid at $\mu=\mu_{co}^+(T)$. For small wall-fluid attraction the local density near the wall can be
depleted so that the adsorption, $\Gamma$, as defined by (\ref{eq:Gammadef}), is negative but finite and one finds
$\gamma_{wl}<\gamma_{wv}+\gamma_{lv}$. This corresponds to the partial drying situation in
Fig.~\ref{fig:contact_angle_schem}, where $\cos(\theta)>-1$. On reducing $\epsilon_w$ further, to a value
$\epsilon_{wd}$, a drying transition can occur whereby for $\epsilon\leq\epsilon_{wd}, \Gamma=-\infty$, and
$\gamma_{wl}=\gamma_{wv}+\gamma_{lv}$ , i.e. $\cos(\theta)=-1$, consistent with the wall-liquid interface being wet by a
macroscopically thick layer of vapor.

We focus on drying and suppose there is a critical drying transition. The divergence of the adsorption is described by

\be
|\Gamma|\sim (\delta\epsilon_w)^{-\beta_s}, \hspace*{5mm}\delta\epsilon_w\to 0
\label{eq:modGam}
\ee
with $\delta\epsilon_w\equiv\epsilon_w-\epsilon_{wd}$, which is accompanied by a divergence of the parallel (transverse) correlation length $\xi_\parallel$: 

\be
\xi_\parallel\sim(\delta\epsilon_w)^{-\nu_\parallel}, \hspace*{5mm}\delta\epsilon_w\to 0
\label{eq:xipardiv}
\ee
 i.e.~density fluctuations parallel to the wall become long-ranged on approaching the transition. This is most easily
understood in terms of the surface structure factor introduced in (\ref{eq:Sq}). One expects Ornstein-Zernike behaviour:
$S(z;q)=S(z;0)/(1+\xi_\parallel^2q^2)$, for small wave numbers $q$, when $z$ is located close to $l$, the thickness of
the vapor film. The latter is given by $l=-\Gamma/(\rho_l-\rho_g)$. Moreover, considerations of capillary wave
fluctuations in the emerging liquid-vapor interface lead to the prediction \cite{Evans:1989aa}:
 
 \be
 \chi(z)=\beta S(z;0)\sim\rho^\prime(z)\xi_\parallel^2, \hspace*{5mm} z\approx l
 \label{eq:chizS}
 \ee
Note that $\rho^\prime(l)$ is the gradient of the density profile of the emerging gas-liquid interface, as $l\to\infty$. Capillary wave arguments then predict  $\rho^\prime(l)\sim\xi_\perp^{-1}$ as $\delta\epsilon_w\to 0$  where $\xi_\perp$ is the width of the depinning gas-liquid interface, i.e. the interfacial roughness. In spatial dimension $d =3$ one has $\xi_\perp^2\sim(2\pi\beta\gamma_{lv})^{-1}\ln(\xi_\parallel/\xi_b$), where $\xi_b$ is the bulk correlation length of the phase that wets; in the present case the vapor. Using the relation (\ref{eq:chiex}), i.e. integrating directly (\ref{eq:chizS}), we expect the surface excess compressibility to diverge as 
\be
\chi_{ex}\sim\xi_\parallel^2\sim(\delta\epsilon_w)^{-2\nu_\parallel}\:.
\label{eq:chiex2}
\ee
It is also important to consider the quantity  $\chi_1\equiv (\partial\Theta/\partial\mu)_T$ introduced in (\ref{eq:maxwell}). Clearly this corresponds to an integral of $\chi(z)$ weighted with the wall-fluid potential; see (\ref{eq:Theta}). The Maxwell relation (\ref{eq:maxwell}) with (\ref{eq:modGam}) dictates that 

\be
\chi_1\sim(\delta\epsilon_w)^{-\beta_s-1}    , \hspace*{5mm}\delta\epsilon_w\to 0                                    
\ee
The other key exponent, $\alpha_s$, is associated with the singular part of the surface tension $\gamma=\Omega_{ex}/A$. For drying one has  $\gamma_{wl}\equiv\gamma_{wv}+\gamma_{lv}+\gamma^{sing}$ with

 \be
|\gamma^{sing}|\sim (\delta\epsilon_w)^{2-\alpha_s}                                                          
 \label{eq:gammasing}
 \ee
Note that  from Young's equation (\ref{eq:Young}) it follows that $\gamma^{sing}=-\gamma_{lv}(1+\cos(\theta)$; this quantity is negative in the partial drying regime. The three critical exponents are not independent. A scaling hypothesis for $\Omega_{ex}$ and a thermodynamic argument employing (\ref{eq:dOmega})  \cite{Evans:1989aa}  both yield the analogue of the well-known Rushbrooke exponent (in)equality for the corresponding bulk exponents, i.e.
\be
2-\alpha_s=2(\nu_\parallel-\beta_s)\:.
\label{eq:rushbrooke}
\ee
       
The critical exponents depend on the dimensionality $d$ and on the ranges of the fluid-fluid ($ff$) and wall-fluid ($wf$)
potential, e.g \cite{Dietrich:1988et}. We assume that the hyperscaling relation $2-\alpha_s=(d-1)\nu_\parallel$ is valid
for the $d-1$ dimensional interface of the $d$ dimensional fluid for $d\leq d_c$, the upper critical dimension. If $ff$
and $wf$ potentials are both of finite range, as in the case of a truncated LJ fluid (\ref{eq:Simpot}) near a square-well
wall (\ref{eq:SRpot}), mean-field (MF) analysis, e.g.~\cite{Dietrich:1988et} yields $\alpha_s=0,\beta_s=0$ (logarithmic
divergence) and $\nu_\parallel=1$. These MF results for SR potentials are consistent with (\ref{eq:rushbrooke}) and
when inserted into the hyperscaling relation yield $d_c=3$. As outlined in the Introduction, RG calculations, Ising
model simulations and concerted theoretical effort to incorporate relevant fluctuation effects conclude that the
critical exponents $\nu_\parallel$ and $\alpha_s$ should depend on the dimensionless parameter
$\omega=(4\pi\beta\gamma_{lv}\xi_b^2)^{-1}$ that measures the strength of interfacial fluctuations. 
Mean field corresponds to an infinitely stiff interface: $\omega=0$.

By contrast, when {\em both} $ff$ and $wf$ potentials exhibit appropriate algebraic (power-law) decay critical drying can occur
and one finds, using hyperscaling, that the upper critical dimension $d_c<3$. Thus MF results for critical exponents should
remain valid in $d=3$ when both potentials are LR. An explicit DFT calculation for such a situation is described in
\cite{Stewart:2005qd}. In the present paper we focus primarily on the case of a truncated $ff$ potential and a $wf$ potential
with attraction that decays algebraically, as $\sim z^{-3}$, see (\ref{eq:LRpot}). For this special case, a MF binding potential analysis (see
Sec.\ref{sec:bp}) yields MF exponents which i) are different from those pertinent to SR potentials described above, ii) satisfy
(\ref{eq:rushbrooke}) and iii) when inserted into the hyperscaling relation, imply $d_c=3$. The same calculation finds that (critical)
drying occurs in the limit $\epsilon_w\to 0$, i.e. where the $wf$ attraction is vanishing and the (microscopic) DFT calculations presented
in Sec.~\ref{sec:dftresults} confirm this result. A simple RG treatment of fluctuation effects, Sec.~\ref{sec:rg}, finds that the critical exponents
are unchanged from their MF values.

\subsection{Phase diagram of a fluid in a slit pore with identical walls}
\label{sec:pdslit}

\begin{figure}[h]
\includegraphics[type=pdf,ext=.pdf,read=.pdf,width=0.94\columnwidth,clip=true]{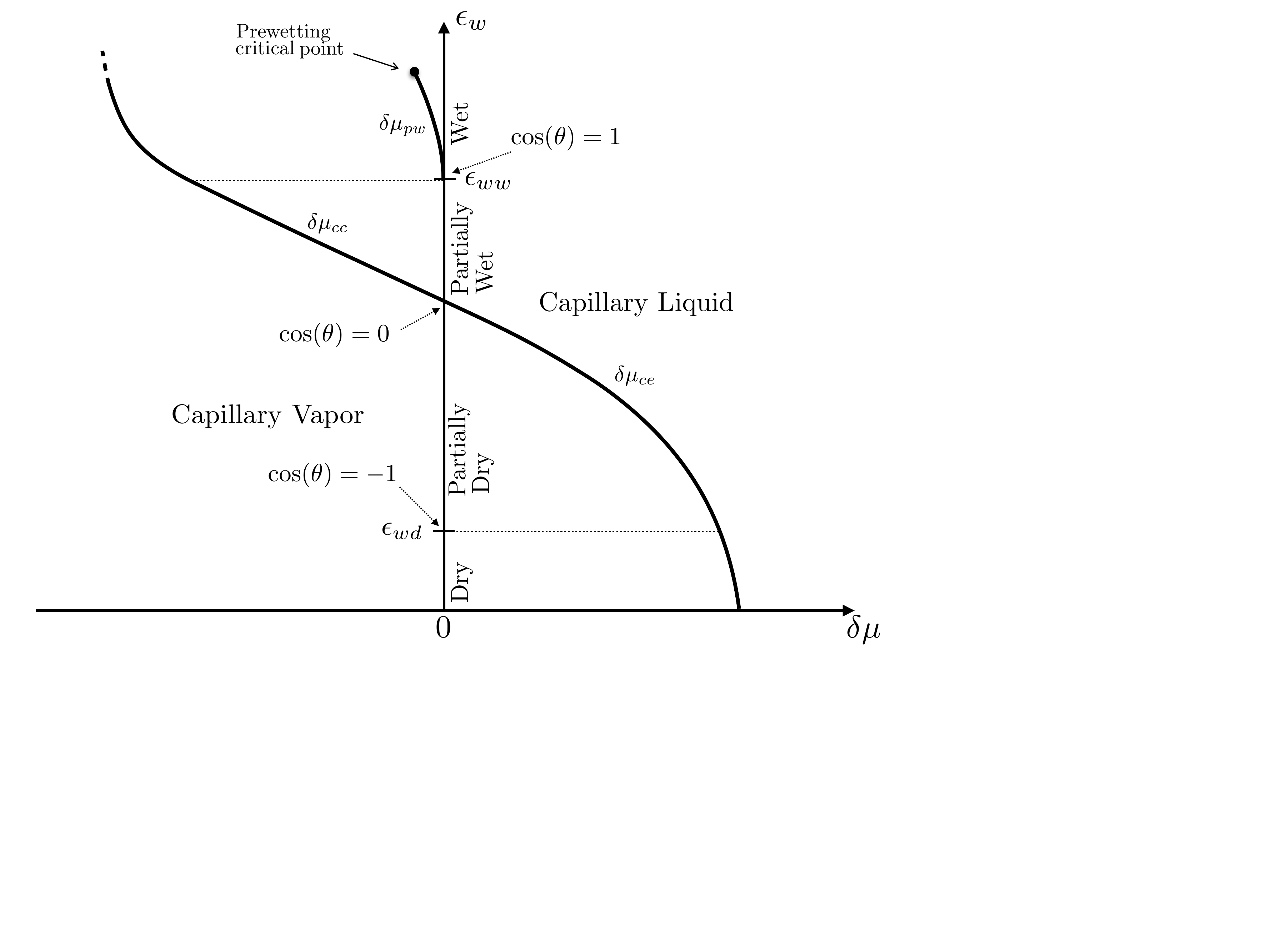}

\caption{Schematic surface phase diagram for a fluid in a slit pore at some $T<T_c$, showing a first order wetting
transition at $\epsilon_{ww}$ and a continuous drying transition at $\epsilon_{wd}$. The situation depicted here is for
wall separation $D$ where capillary condensation occurs for $\delta\mu_{cc}\ll\delta\mu_{pw}$. Prewetting ($pw$) is then
metastable. The $cc$ line can, for suitable $D$, cross the prewetting line giving a prewetting triple point
\cite{Evans:1985ss}. For a critical drying transition the $ce$ line enters almost vertically into $\epsilon_{wd}$, but
not precisely vertical because of H.O. corrections to (\ref{eq:kelvin}). (The corrections involve powers of $l/D$ in the
wetting and drying regimes; $l$ is the film thickness.) }

\label{fig:schematicpd}
\end{figure}

In order to set the scene and assist the reader, we outline here general
features of the surface phase behaviour of a simple fluid in a slit
pore having wall separation $D$, which is the system that we consider in the present simulations. The natural
choice of variables for representing the phase behaviour is the pair
of fields: $\delta\mu\equiv\mu-\mu_{co}$, which measures the deviation
of the chemical potential from its bulk coexistence value; and
$\epsilon_{w}$, which measures the strength of the wall-fluid
attraction. Fig.~\ref{fig:schematicpd} shows a schematic sketch of one
possible phase diagram for such a system at some temperature $T<T_c$,
with $T_c$ the bulk critical temperature. 

The system exhibits two lines of phase coexistence: the capillary line
and the prewetting line. The capillary line is the locus of state
points for which gas-liquid coexistence occurs in the slit pore. It is given approximately by the Kelvin equation 
\be
\delta\mu_{ce} \approx\frac{-2\gamma_{lv}\cos(\theta)}{D(\rho_l-\rho_v)}
\label{eq:kelvin}
\ee
where $\rho_l$ and $\rho_v$ are the coexisting liquid and vapor densities, respectively. Small values of $\epsilon_{w}$
favor the capillary gas phase while large values favor the capillary liquid. If the transition occurs for $\delta\mu<0$
it is referred to as capillary condensation ($cc$); if the transition occurs for $\delta\mu>0$ it referred to as
capillary evaporation ($ce$). The capillary condensation/evaporation line is sketched within a range of $\epsilon_{w}$
between the drying wall strength $\epsilon_{wd}$ at the lower end and the wetting wall strength $\epsilon_{ww}$ at the
upper end. The line in Fig.~\ref{fig:schematicpd} extends smoothly to $\epsilon_w<\epsilon_{wd}$ and to
$\epsilon_w>\epsilon_{ww}$, as shown. The detailed behaviour depends on higher order contributions in (\ref{eq:kelvin}).
At $\epsilon_w\geq\epsilon_{ww}$ the contact angle is zero, i.e. $\cos(\theta)=1$, while at $\epsilon\leq \epsilon_{wd}$
it is $180^\circ$ i.e. $\cos(\theta)=-1$. The neutral wall, at which $\cos(\theta)=0$, corresponds to the value of
$\epsilon_{w}$ at which the capillary line crosses the $\delta\mu=0$ axis. In the sketch in Fig~\ref{fig:schematicpd} we
have drawn the situation that occurs for critical drying and first order wetting, see Sec~\ref{sec:bp}. This scenario
pertains to the LR wall-fluid potential that we investigate here, though for the particular model that we consider
$\epsilon_{wd}=0$. Note however, that other scenarios are possible. Specifically, and as we show below, wetting can
also be continuous (critical). The prewetting ($pw$) line shown occurs only for first order wetting. It emerges
tangentially from the wetting point $\delta\mu=0^-,\epsilon_w=\epsilon_{ww}$ and extends some distance to $\delta\mu<0$
before terminating at a prewetting critical point whose critical properties correspond to the universality class of the 2d
Ising model~\cite{Nicolaides:1989aa}. Along the prewetting line a thin layer of liquid on each wall coexists with a
thick but finite liquid layer \cite{Dietrich:1988et,Fan:1993aa}. However we will not consider prewetting in the present
work.

Eq.~(\ref{eq:kelvin}) proves a useful estimate of the range of $\beta\delta\mu$ for which capillary condensation and evaporation can be found in our system. At drying or wetting (\ref{eq:kelvin}) yields $-\delta\mu_{cc}^w=\delta\mu_{ce}^d=2\gamma_{lv}/D(\rho_l-\rho_v)$ and the horizontal scale in Fig~\ref{fig:schematicpd} is determined by
$\beta\mu_{ce}^d\approx 2\beta\gamma_{lv}\sigma^2/[D\sigma^{-1}(\rho_l\sigma^{3}-\rho_v\sigma^3)]$. For our simulated LJ fluid at $T/T_c=0.775$ the liquid-vapor surface tension is given by $\beta\gamma_{lv}\sigma^2=0.404$ and $\beta\mu_{ce}^d=1.295/D\sigma^{-1}$. Thus for $D=30\sigma$, the wall separation in most of the simulations, we find $\beta\mu_{ce}^d\approx 0.04$. The physics we describe is occurring at small under or over saturations. But these values of $\beta\delta\mu$ are certainly pertinent to experiment.

\section{Theory for a Model Fluid with SR Fluid-Fluid and LR Wall-Fluid Potentials} 
\label{sec:theory}
\subsection{Binding potential analysis}
\label{sec:bp}

We follow the standard treatment, e.g.~\cite{Dietrich:1988et}, of wetting/drying
transitions and consider $\omega^{ex}(l)$, the excess grand potential
per unit surface area, as a function of the thickness $l$ of the
wetting/drying layer. For a truncated LJ model adsorbed at a single
wall exerting the potential (\ref{eq:LRpot}) or the modified version, we expect

\be
\omega^{ex}(l)=\gamma_{wv}+\gamma_{lv}+\omega_B(l)+\delta\mu (\rho_l-\rho_v)l
\label{eq:grandpot}
\ee
with the binding potential 

\be
\omega_B(l)=a\exp{(-l/\xi_b)}+bl^{-2}+{\rm H.O.T.}
\label{eq:bpot}
\ee
As previously, $\rho_l$ and $\rho_v$ are the liquid and vapor densities at
coexistence, $\delta\mu=\mu-\mu_{co}\ge 0$ is the deviation of the
chemical potential from its value at coexistence and we have
specialized now to the case of drying, i.e.  $l$ is the thickness of a
layer of vapor that can intrude between the weakly attractive wall and
the bulk liquid at $z= \infty$. In the limit of complete drying, at
$\delta\mu=0^+$, $l$ diverges and the $wl$ interface is a composite of
the $wv$ and $lv$ interfaces. In this limit
$\gamma_{wl}=\gamma_{wv}+\gamma_{lv}$, i.e. $\cos(\theta)= -1$, as mentioned previously in Sec.~\ref{sec:phenomenology}. The
binding potential in (\ref{eq:bpot}) has two leading contributions. The
exponential term accounts for SR fluid-fluid interactions; $\xi_b$ is
the true correlation length of the bulk phase that wets, in our case
the vapor, and $a$ is a positive coefficient.  The term $bl^{-2}$ is
associated with the $z^{-3}$ decay of $W_{LR}(z)$ in (\ref{eq:LRpot}); it arises from
dispersion (van der Waals) forces between the substrate and the
fluid. The higher order terms in (\ref{eq:bpot}) include higher inverse powers such as $cl^{-3}$ 
as well as more rapidly decaying exponentials. We ignore these
in the subsequent analysis.  Making a straightforward sharp-kink
approximation, or Hamaker type calculation, e.g. ~\cite{Dietrich:1988et}, yields

\be
b=-(\rho_l-\rho_v)\epsilon_w\epsilon_{LJ}\sigma^3/2
\label{eq:b}
\ee
Since $b<0$ for all $T<T_c$, minimizing (\ref{eq:grandpot}) w.r.t. $l$ at $\delta\mu=0^+$, leads to a
finite value for the equilibrium thickness:

\be
\frac{-l_{eq}}{\xi_b}=\ln\epsilon_w-3\ln\left(\frac{l_{eq}}{\xi_b}\right)+{\rm constants};\hspace*{1mm} \delta\mu=0^+
\label{eq:leq}
\ee

A formula equivalent to (\ref{eq:leq}) was derived by Nightingale {\em et
    al.} (see Eq. 6 of \cite{Nightingale:1983ty}) in a
    study of critical wetting in systems with LR forces. Those authors
    considered only the case where $\epsilon_w >0$ and concluded there
    was no wetting, critical or first order. Here we focus on the
    situation where $\epsilon_w \to 0^+$, $b\to 0^-$ and $l_{eq}$ diverges continuously. Note
    that for $\epsilon_w =0$, $W_{LR}(z)$ in (\ref{eq:LRpot}) reduces to the planar hard-wall potential
    and minimization of (\ref{eq:grandpot}) then yields $-l_{eq}/\xi_b=\ln(\delta\mu)+{\rm const}$, the mean-field (MF) result
    appropriate for complete drying from off-coexistence, for all $T<T_c$, e.g.~\cite{Evans:1992jo,Dietrich:1988et,Evans:1990aa}.  

Using (\ref{eq:grandpot},\ref{eq:bpot}) we can calculate several properties
    and examine these, within MF, in the approach to critical drying
    $\epsilon_w\to 0^+$. The local compressibility, evaluated for $z\approx l_{eq}$, is given
    by \cite{Evans:2015aa,Evans:1990aa}

\be
\chi(l_{eq})=\left(\frac{\partial\rho(z)}{\partial\mu}\right)_{z=l_{eq}}\sim-\rho^\prime(l_{eq})\left(\frac{\partial l_{eq}}{\partial \mu}\right)
\label{eq:Chi}
\ee
where the prime denotes differentiation w.r.t. $z$. From (\ref{eq:grandpot}) it follows that, at leading order, 

\be
\left(\frac{\partial l_{eq}}{\partial\mu}\right)=-\frac{\xi_b^2}{a}(\rho_l-\rho_v)\exp({l_{eq}/\xi_b}); \hspace*{1mm}\delta\mu=0^+
\label{eq:leqder}
\ee
Capillary wave arguments predict that in the limit of critical drying
$\rho^\prime(l_{eq})\sim\xi_\perp^{-1}$, where $\xi_\perp$ is the
interfacial roughness introduced above. Within MF, $\xi_\perp^{-1}$ is non-zero, and using
(\ref{eq:leq}) we deduce 

\be
\ln\chi(l_{eq})\sim \frac{l_{eq}}{\xi_b} + {\rm const.};\hspace*{2mm} \delta\mu=0^+
\label{eq:logchi}
\ee
The predictions (\ref{eq:leq}) and (\ref{eq:logchi}) were tested carefully using the microscopic DFT, as described below.
The quantity $\Gamma_1$ defined in (\ref{eq:maxwell}), is proportional to $-\partial l_{eq}/\partial\epsilon_w$. It follows that 
as $\epsilon_w\to 0^+$,

\be
\chi_1=\Gamma_1\sim\epsilon_w^{-1}(1-3(\ln\epsilon_w)^{-1}); \hspace*{2mm}\delta\mu=0^+
\ee

We can also extract the correlation length $\xi_\parallel$ that
describes density-density correlations parallel to the wall. General
arguments predict that
$\xi_\parallel^2$ diverges in the same way as the surface excess
compressibility, see (\ref{eq:chiex2}).

Since $\chi_{ex}$ is proportional to $-\partial l_{eq}/\partial\mu$ it follows
from (\ref{eq:leqder},\ref{eq:leq}) that $\xi_\parallel$ diverges as

\be
\xi_\parallel\sim\epsilon_w^{-1/2}(-\ln\epsilon_w)^{3/2}; \hspace*{1mm}\delta\mu=0^+
\label{eq:xipar}
\ee 
in the limit $\epsilon_w \to 0^+$. The same result is obtained
from standard binding potential considerations \cite{Dietrich:1988et} where one has
$\xi_\parallel^{-2}\propto \partial^2\omega_B(l)/\partial l^2$ at
$l=l_{eq}$.

The variation of $\cos(\theta)$ close to critical
drying is determined by $\omega_B(l_{eq})$ at $\delta\mu=0^+$, i.e. the
singular part of the surface excess free energy $\gamma^{sing}$. Using Young's
equation (\ref{eq:Young}) one finds $1+ \cos(\theta)=-\omega_B(l_{eq})/\gamma^{lv}$ and for the present binding potential (\ref{eq:bpot})
we obtain

\be
1+\cos(\theta)\sim\epsilon_w(-\ln\epsilon_w)^{-2}
\label{eq:costheta}
\ee
in the limit $\epsilon_w\to 0^+$. This result is striking. Were the
logarithm not present in (\ref{eq:costheta}) the theory would predict $1+ \cos(\theta)$
vanishing linearly with $\epsilon_w$, a signature of a 1st order drying
transition. It is only the presence of the logarithm that ensures a
continuous (critical) transition. The critical exponent $\alpha_s$,
defined by the vanishing of the singular part of the surface excess
free energy $|\gamma^{sing}|\sim \epsilon_w^{2-\alpha_s}$, see (\ref{eq:gammasing}), clearly takes the value $\alpha_s =1$, with log
corrections, in this particular case. The situation is similar to that
for complete drying from off-coexistence where for a planar hard-wall,
say, $\gamma^{sing}\sim\delta\mu\ln\delta\mu$, $\delta\mu\to 0^+$. 

It is important to distinguish the MF scenario presented above from
that corresponding to a SR wall-fluid potential such as (\ref{eq:SRpot}). In the SR
case it is well-known, e.g.~\cite{Brezin:1983dn,Dietrich:1988et}, that the second inverse power-law
term in (\ref{eq:bpot}) must be replaced by a H.O. term proportional
to $\exp(-2l/\xi_b)$ while the coefficient of the leading
$\exp{(-l/\xi_b)}$ term now depends on $\epsilon_w$:
$a(\epsilon_w)\sim (\epsilon_w- \epsilon^{MF}_{wc})$ where
$\epsilon^{MF}_{wc}>0$ is the strength of the wall-fluid attraction at
which critical drying occurs in MF. Defining
$\delta\epsilon_w=\epsilon_w- \epsilon^{MF}_{wc}$, MF analysis for the SR
case yields, for $\delta\mu=0^+$,

\bea
\frac{-l_{eq}}{\xi_b}&\sim&\ln(\delta\epsilon_w),\label{eq:SRleqa}\\
\chi(l_{eq})&\sim& (\delta\epsilon_w)^{-2},\\ 
\xi_\parallel&\sim& (\delta\epsilon_w)^{-1},\\
\chi_1&\sim&(\delta\epsilon_w)^{-1},
\label{eq:SRleq}
\eea

and

\be
1+\cos(\theta)\sim(\delta\epsilon_w)^2 \hspace*{2mm} {\rm or} \hspace*{2mm} \alpha_s=0;
\label{eq:SRcostheta}
\ee
The critical exponents $\beta_s,\nu_\parallel$ and $\alpha_s$ take the values mentioned in Sec.~\ref{sec:phenomenology}.
These results are clearly very different from those we obtained above for the LR case.

It is also important to consider the implications of a binding potential of the form (\ref{eq:bpot}) for wetting. In this context we recall an argument of Ebner and Saam (ES) \cite{Ebner:1987xy} who considered a lattice gas (Ising) model for which the substrate-fluid potential is $W_n=-RJn^{-p}$, $p\geq 3$, where $n$ labels the $n$th layer from the substrate and $RJ>0$ is a constant. For the case of a SR $ff$ potential the most slowly decaying term in the ES grand potential, i.e. the binding potential, has the form $(\rho_\alpha-\rho_\beta)RJ/(\rho-1)l^{p-1}$ where $\rho_\alpha$ is the density and $l$ is the thickness of the phase $\alpha$ that wets the substrate. In the case of wetting by the denser `liquid' $\rho_\alpha>\rho_\beta$, the density of the `vapor', and this term $>0$, implying the binding potential can have a relative minimum at $l=\infty$. Suppose that there is incomplete wetting, for some value of $RJ$. Then ES argue the binding potential must have a minimum, lower than at $l=\infty$, for a `liquid' film of finite thickness. On increasing $RJ$, equivalent to increasing $\epsilon_w$ in our system, a wetting transition can occur but this can only be first order: there cannot be a continuous evolution from finite $l$ to infinite $l$. Thus if a wetting transition occurs this cannot be critical. For sufficiently large $RJ$, or $\epsilon_w$, one expect on physical grounds that wetting should occur. It follows this must be first order.

ES also consider drying where $\alpha$ now corresponds to `vapor' and $\beta$ to `liquid'. Now the relevant term in the binding potential is negative, as given by (\ref{eq:b}), and there is a relative maximum at $l=\infty$. ES then argue that drying cannot occur for any $T<T_c$. They do not consider the limit $RJ\to 0$, corresponding to our present limit $\epsilon_w\to 0^+$.

The numerical work of ES for the lattice gas model confirms that wetting is always first order and ES find no critical or first order drying transitions. In our present DFT and simulation studies we find that for $W_{LR}(z)$ in (\ref{eq:LRpot}) the wetting transition is first order. In contrast to ES, we do find a critical drying transition. This occurs as the attractive strength $\epsilon_w\to0^+$. In this limit our wall-fluid potential reduces to that of a hard-wall, for which drying occurs for all $T<T_c$. This hard-wall boundary condition, particular to fluids, drives the drying transition. Both DFT and simulation find critical drying as $\epsilon_w\to 0^+$.

\subsection{Renormalization Group (RG) treatment of fluctuations}
\label{sec:rg}
The analysis described in Sec.~\ref{sec:bp} was strictly MF; this omits some of the effects of capillary wave (CW) fluctuations. For example, for infinite surface area, MF predicts a sharp interface with $\xi_\perp$ finite in all dimensions $d$ whereas, in reality, we expect $\xi_\perp$ to diverge for $d\le 3$. An important early attempt to incorporate CW fluctuations was that of Brezin et al. \cite{Brezin:1983dn} who introduced a RG treatment for the case of SR forces where the upper critical dimension $d_c=3$ for both critical wetting and complete wetting from off-coexistence. We follow their methodology for our binding potential (\ref{eq:bpot}).

First we invoke the hyperscaling relation $(2-\alpha_s) =(d-1)\nu_\parallel$, where $\nu_\parallel$ is the critical exponent for $\xi_\parallel$, insert the MF exponents given in Sec.~\ref{sec:bp}, and deduce that the upper critical dimension is $d_c =3$ for the present system. Introducing again the standard, dimensionless parameter   $\omega$, that measures the strength of CW fluctuations, the RG treatment then implies we should consider an effective binding potential (renormalized) at the scale $\xi_\parallel$:
\be
\omega_{\xi_\parallel}(l)=a \xi_\parallel^\omega\exp{(-l/\xi_b)}+bl^{-2}+\delta\mu(\rho_l-\rho_v)l
\label{eq:omegaxi}
\ee
The exponential term is renormalized but the remaining power-law terms are not; in particular the coefficient $b$ is assumed to be unchanged. Minimization of (\ref{eq:omegaxi}) yields 

\be
-\frac{l_{eq}}{\xi_b}=(1+\frac{\omega}{2})(\ln\epsilon_w-3\ln(l_{eq}/\xi_b)); \hspace*{2mm}\delta\mu=0^+
\label{eq:leqrg}
\ee
as $\epsilon_w\to 0^+$. The equilibrium thickness still diverges with
the MF form (\ref{eq:leq}) but the amplitude is increased by a factor
$(1+\omega/2)$. MF is recovered when the interface becomes very stiff
so that $\omega\to 0$. The parallel correlation length can be obtained
from either $\xi_\parallel^{-2}\propto
\left(\frac{\partial^2\omega_B(l)}{\partial l^2}\right)$ at $l=l_{eq}$
or from $\xi_\parallel^2\propto \left(\frac{\partial
  l_{eq}}{\partial\mu}\right)$, see (\ref{eq:chiex2}). In both cases we find as $\epsilon_w \to 0^+$

\be
\xi_\parallel\sim\epsilon_w^{-1/2}[(1+\frac{w}{2})(-\ln\epsilon_w)]^{3/2};\hspace*{2mm}\delta\mu=0^+
\ee

The singular part of the surface excess free energy can be calculated from (\ref{eq:omegaxi}) and we obtain 
\be
1+\cos(\theta)\sim\epsilon_w(-(1+\frac{\omega}{2})\ln\epsilon_w)^{-2}; \hspace*{2mm}\delta\mu=0^+
\ee
Once again only the amplitudes are changed from the MF results (\ref{eq:xipar}) and (\ref{eq:costheta}). Note that (\ref{eq:leqrg}) is reminiscent of the result for complete drying from off-coexistence for SR forces, e.g. at a planar hard-wall. There the second term in the r.h.s. of (\ref{eq:omegaxi}) is absent but the third remains leading to 
\be
-\frac{l_{eq}}{\xi_b}=(1+\frac{\omega}{2}) \ln\delta\mu,  {\rm as } ~\delta\mu\to 0^+
\ee                                      

Unlike the case of SR forces considered by Brezin et
al.~\cite{Brezin:1983dn} and in many subsequent studies,
e.g. \cite{Binder:1989aa,Albano:2012db,Parry:2008ef,Parry:2008rp,Fisher:1985aa,Parry:2009aa}
where several of the critical exponents for critical wetting are predicted to depend
explicitly on the parameter $\omega$, for the binding potential
(\ref{eq:bpot}) our RG analysis predicts the critical exponents to be
unchanged from their MF values and therefore independent of $\omega$
even though the upper critical dimension is also $d_c=3$. We note that
the conclusions of the MF and RG analyzes are changed little if we
consider LR wall-fluid potentials other than the standard $9$-$3$ case
(\ref{eq:LRpot}). Suppose the leading wall-fluid power-law decay is proportional
to $-(\sigma/z)^p$, with $p>2$. Then the coefficient of the second term
in (\ref{eq:leq}) is replaced by $p$, (\ref{eq:logchi}) is
unchanged, and the power of the logarithm in (\ref{eq:xipar}) and
(\ref{eq:costheta}) is replaced by $p/2$ and $-(p-1)$, respectively. 
The RG results are changed accordingly.

\subsection{DFT Treatment}

\label{sec:DFTmethod}

The classical DFT that we employ is that used in a previous study of solvophobic substrates but one that did not address
critical drying \cite{Evans:2015aa}. The excess Helmholtz free energy functional is approximated by the sum of a
hard-sphere functional, treated by means of Rosenfeld's fundamental measure theory, and a standard MF treatment of
attractive fluid-fluid interactions. Eq.~(14) of Ref.~\cite{Evans:2015aa} displays the grand potential functional. This
form of the functional has been used in many studies of fluid interfacial phenomena, including wetting and capillary
confinement; Ref \cite{Evans:2015aa} provides pertinent references. In the present study the attractive part of the
truncated LJ potential is given by

\be
\phi_{\rm att}(r)=\left \{ \begin{array}{ll}
-\epsilon_{LJ}, \mbox{\hspace{4mm}}   &  r<r_{\rm min}  \\
 4\epsilon_{LJ}\left[\left(\frac{\sigma}{r}\right)^{12}-\left(\frac{\sigma}{r}\right)^{6}\right], & r_{\rm min}<r<r_c \, ,\\
0, & r>r_c,\\
\end{array}
\right.
\label{eq:DFTpot}
\ee
where $r_{\rm min}=2^{1/6}\sigma$. The potential is truncated at $r_c
=2.5 \sigma$, as in simulation. The critical temperature is given by
$k_BT_c=1.3194\epsilon_{LJ}$ and calculations are performed at $T
=0.775T_c$. The LR wall-fluid potential is the standard $9$-$3$ model given by $W_{LR}(z)$ in (\ref{eq:LRpot}).
We also investigate the SR case (\ref{eq:SRpot}). The hard-sphere diameter, entering the hard sphere functional, is $d=\sigma$.

In the DFT calculations we determine equilibrium density profiles
$\rho(z)$ and the surface tensions
$\gamma_{lv},\gamma_{wl},\gamma_{wv}$, by minimizing the grand
potential functional \cite{Evans:2015aa}. The local compressibility
$\chi(z)$ defined in (\ref{eq:chizdef}) is determined numerically as described in
Ref. \cite{Evans:2015aa}. We have performed calculations for a single wall
and for a pair of confining walls, equivalent to the GCMC
simulations. In Sec.~\ref{sec:dftsingle} we show results for the single wall and in Sec.~\ref{sec:twowalls} for two walls.

\section{Results from DFT}
\label{sec:dftresults}

\subsection{Fluid adsorbed at a single planar wall}

\label{sec:dftsingle}

\begin{figure}[h]
\centerline{\includegraphics[width=8cm,clip=true]{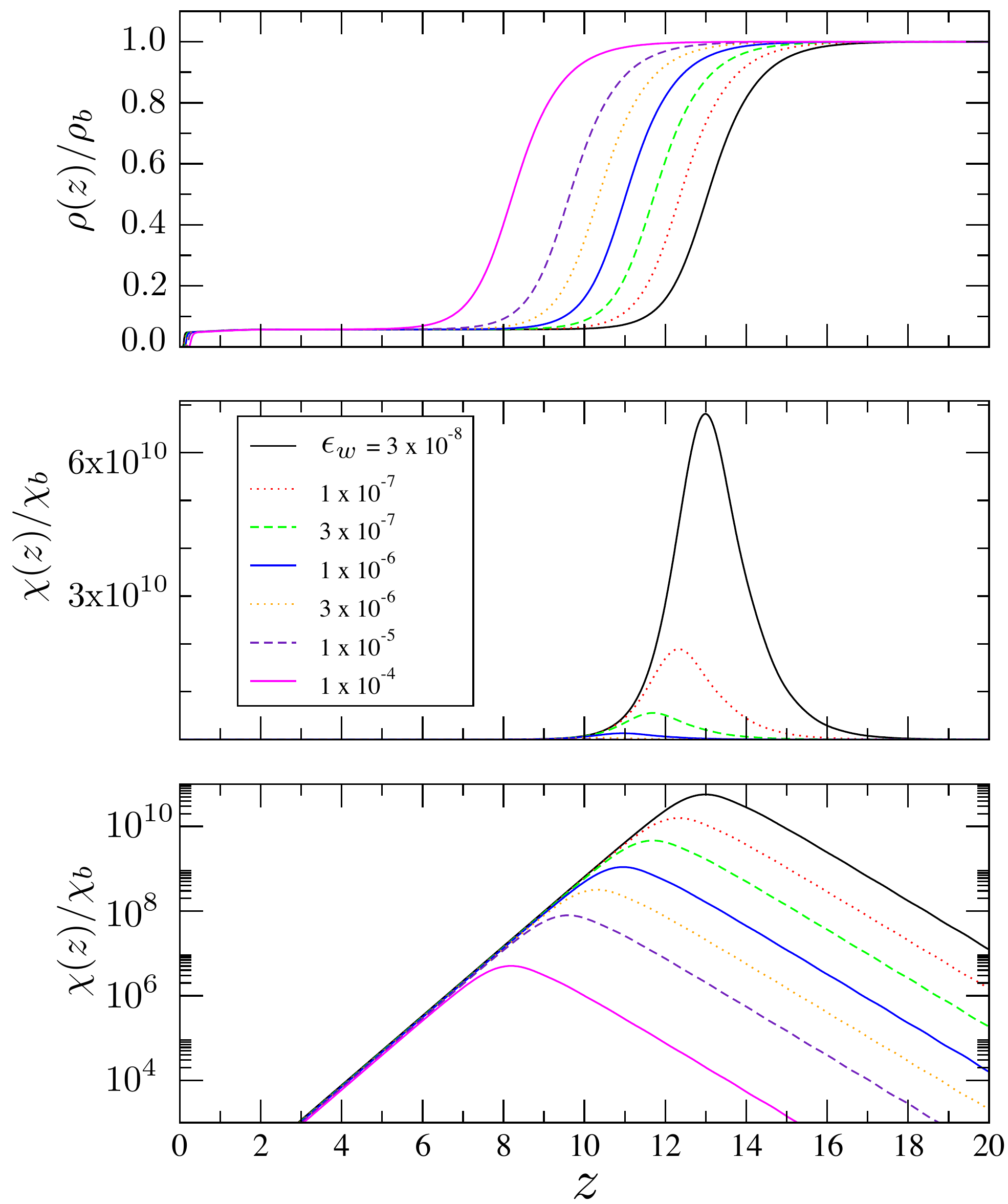}}
\caption{DFT results for the normalised density profiles $\rho(z)/\rho_b$
  (top panel) and the local compressibilities $\chi(z)/\chi_b$ (linear scale
  - middle panel; log-scale - bottom panel) for the fluid at a single
  LR wall. The strength of the wall-fluid interaction potential $\epsilon_w$ is
  given in the key. The temperature is $T = 0.775T_c$ and the
  reservoir is at bulk liquid-gas coexistence, on the liquid side,
  $\delta\mu = 0^+$. }
\label{fig:mariafig}
\end{figure}

\begin{figure}[h]
\includegraphics[type=pdf,ext=.pdf,read=.pdf,width=0.94\columnwidth,clip=true]{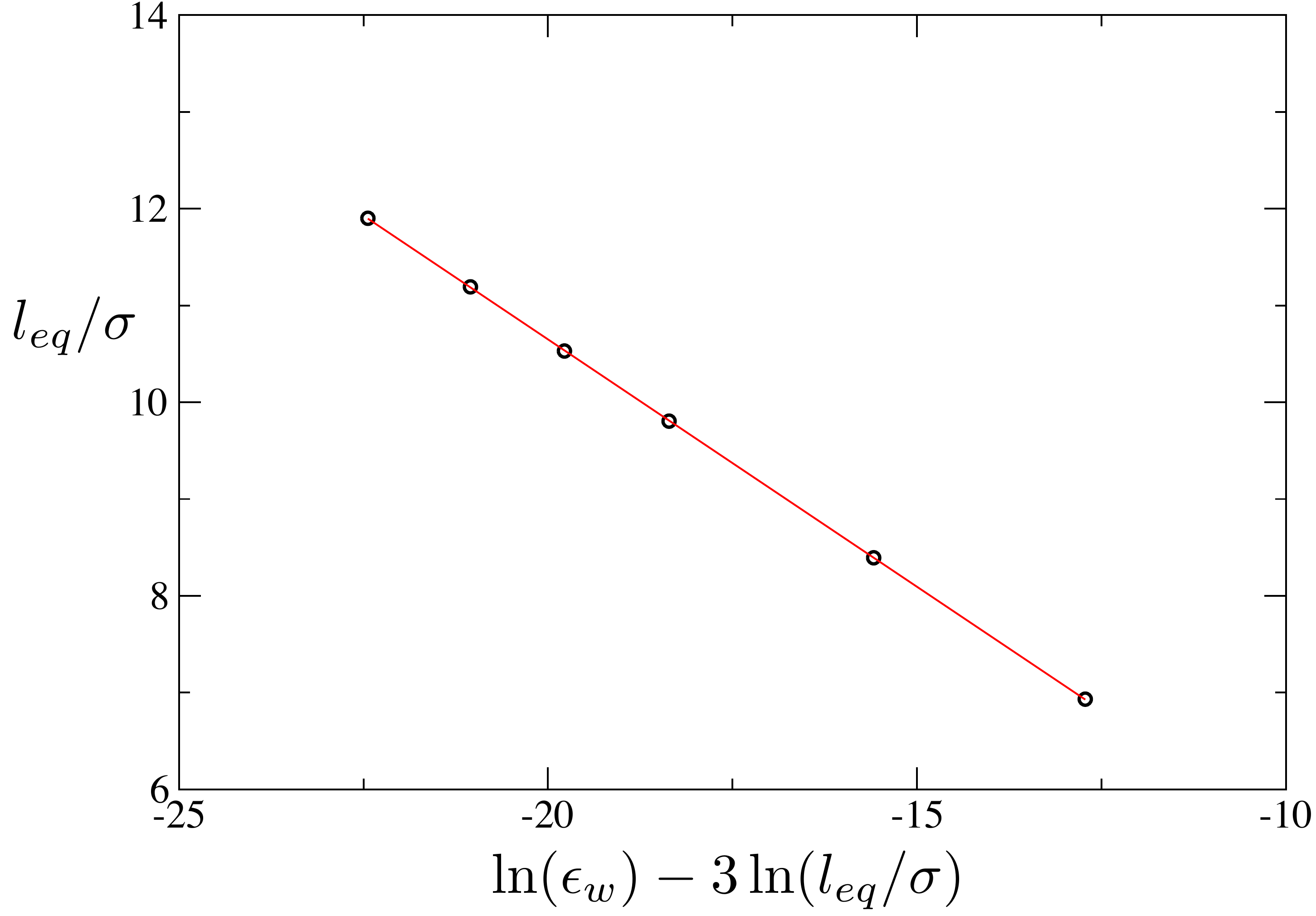}
\caption{DFT results for the equilibrium drying film thickness at various LR wall potentials
 $3 \times 10^{-7} < \epsilon_w < 1 \times 10^{-4}$. $\delta \mu = 0^+$ and the temperature is $T = 0.775T_c$. 
The straight line fit confirms the prediction (\ref{eq:leq}) from the binding potential. }

\label{fig:costhetacompare}
\end{figure}

\begin{figure}[h]
\includegraphics[type=pdf,ext=.pdf,read=.pdf,width=0.94\columnwidth,clip=true]{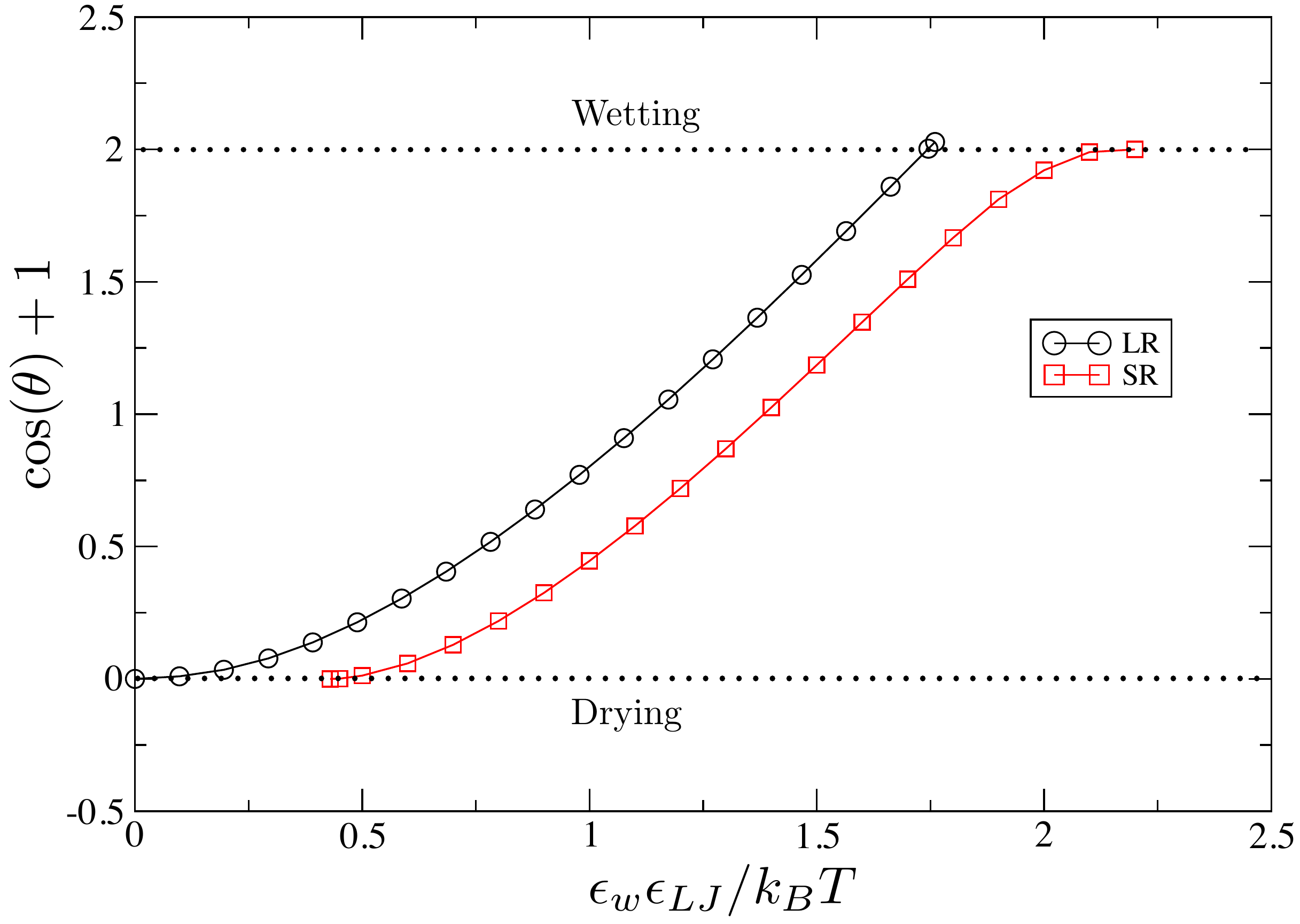}
\caption{ DFT results for $\cos(\theta)+1$ versus  scaled wall-fluid attractive strength (see text)
     for the SR square well potential (\ref{eq:SRpot}) and LR wall potential (\ref{eq:LRpot})  at $T=0.775T_c$. 
Note that for  SR  both wetting and drying are critical while for LR wetting is first order and drying is critical with the transition at $\epsilon_w=0$.}

\label{fig:costhetadft}
\end{figure}

Our key results are shown in Fig.~\ref{fig:mariafig}. Here we plot $\rho(z)$ and $\chi(z)$ for very small values of
$\epsilon_w$ at the LR wall (\ref{eq:LRpot}). As $\epsilon_w$ is reduced towards zero the thickness of the drying film
$l_{eq}$ increases (top). We have confirmed in detail, within DFT, that the Gibbs adsorption $\Gamma$ or $l_{eq}$ grows
according to (\ref{eq:leq}). This is illustrated in Fig.~\ref{fig:costhetacompare} where we plot $l_{eq}$ vs
$\ln\epsilon_w-3\ln(\l_{eq})$. The slope yields $\xi_b=0.51\sigma$ for the correlation length of the bulk (vapor) phase
that wets. This estimate is close to that from a separate DFT calculation of the binding potential. The position of the
peak in $\chi(z)$ shifts with the position of the gas-liquid interface and its height increases very rapidly as
$\epsilon_w \to 0^+$ (middle). The bottom panel shows clearly that $\ln\chi(l_{eq})$ increases linearly with $l_{eq}$.
The prediction (\ref{eq:logchi}), including the correct prefactor, the inverse bulk correlation length, is confirmed by
our DFT calculations. We determine the contact angle via Young's equation and DFT results for $\cos(\theta)$ are shown
in Fig.~\ref{fig:costhetadft}. Note that in order to facilitate comparison with subsequent plots of simulation results,
the quantity plotted on the abscissa of Fig.~\ref{fig:costhetadft} is the wall-fluid potential strength in units of $k_BT$, i.e. for the LR
case (\ref{eq:LRpot}) this denotes $\epsilon_w\epsilon_{LJ}/k_BT=0.9779\epsilon_w$ and for the square well 
(\ref{eq:SRpot}) denotes the well-depth in units of $k_BT$. For the LR case
(\ref{eq:LRpot}) we find critical drying at $\epsilon_w =0$ and 1st order wetting at a value of $\epsilon_w$ that is
smaller than in simulation, see later. For the SR case (square-well), both drying and wetting are critical transitions,
as found in simulation. However, as we shall see, the separation in $\epsilon_w$ between wetting and drying in DFT is
smaller than in simulation. The microscopic DFT results for a single LR wall yield the same behavior as those from the
simple binding potential treatment, based on (\ref{eq:bpot}), i.e., for the LR case, the DFT yields the same MF critical
exponents, including any logarithmic ($\ln \epsilon_w$) corrections, as those predicted in Sec.~\ref{sec:bp}. This is
not unexpected: DFT is a MF treatment of fluid interfaces and we expect it to capture the same asymptotic, $|\Gamma|$ or
$l_{eq} \to \infty$, behaviour as the binding potential analysis. The key difference between the two approaches lies in
the fact that DFT incorporates accurately the short distance behaviour of the density profile. In particular our DFT
satisfies exactly the hard-wall sum rule: $k_BT\rho(0^+ ) = p(\mu)$, where $p$ is the pressure of the bulk fluid, that
is important in ensuring that complete drying occurs in the limit $\epsilon_w \to 0^+$. This is mimicked in the binding
potential by the first, repulsive, term in (\ref{eq:bpot}).

In our DFT calculations we can compute accurately the excess grand potential $\omega^{ex} (\Gamma)$, for non-equilibrium values
of the adsorption $\Gamma$, during the minimization of the functional. This is equivalent to determining numerically the
binding potential entering (\ref{eq:grandpot}). For the LR case $\omega^{ex} (\Gamma)$ exhibits two minima, corresponding to a microscopic liquid film
and an infinitely thick liquid film, on approaching the wetting transition. At the transition the two minima are equal
but there remains a maximum between these -- a clear signature that the transition is first order and in keeping with the
lattice gas results of ES \cite{Ebner:1987xy}. On the other hand, on approaching the drying transition $\omega^{ex} (\Gamma)$ exhibits a single minimum
at $| \Gamma |$ corresponding to a thick drying film. The minimum erodes continuously and shifts to larger $| \Gamma |$ as $\epsilon_w$ is
reduced. In the limit $\epsilon_w \to 0^+$ the minimum is at $| \Gamma |= \infty$ , i.e. the transition is critical as predicted by the binding
potential treatment. For the SR case $\omega^{ex} (\Gamma)$ exhibits a single minimum in the approach to both wetting and drying
showing that both are critical transitions. We have not attempted to determine critical exponents numerically for the SR case because of the difficulty of locating accurately the drying and wetting points. However, there is no reason to expect the exponents from DFT to differ from those given by the MF analysis of the binding potential, i.e. (\ref{eq:SRleqa})-(\ref{eq:SRcostheta}).

 \begin{figure}[h]
\includegraphics[type=pdf,ext=.pdf,read=.pdf,width=0.94\columnwidth,clip=true]{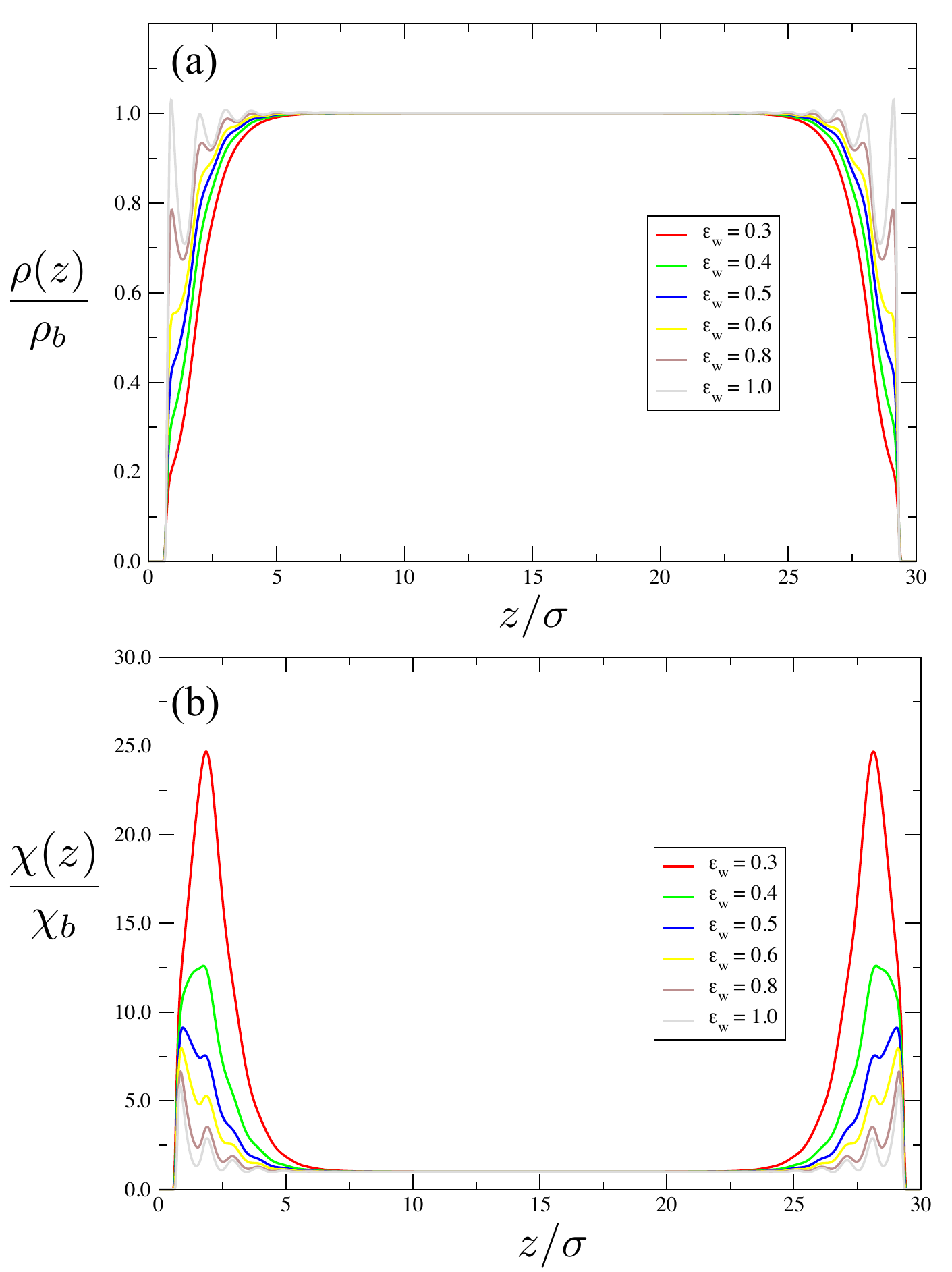}
\caption{{\bf  (a)} DFT results for the density profiles for the liquid confined in a slit for width $D=30\sigma$ for a range of LR wall strengths $\epsilon_w$. 
 $\delta\mu = 0^+$ and $T = 0.775T_c$. {\bf  (b)} Corresponding results for the local compressibility. }

\label{fig:dftslitprofiles}
\end{figure}

\subsection{Fluid adsorbed between two planar walls}
 \label{sec:twowalls}

Since the GCMC simulations, to be described in Secs.~\ref{sec:simmeth} and \ref{sec:simresults}, investigate fluids confined between two planar walls, we
also performed some DFT calculations for the confined system. We focused on the LR case, with each wall described by the
potential (\ref{eq:LRpot}). The wall-separation was chosen to be $D=30\sigma$ and the temperature was $T=0.775T_c$ which correspond to the
system studied in most detail in the simulations. Results for the density profiles and the local compressibility are
shown in Fig.~\ref{fig:dftslitprofiles} for a range of values of $\epsilon_w$. Note that the smallest value is well-removed from the drying point $\epsilon_w = 0^+$;
for $\epsilon_w =0.3$ we find the contact angle $\theta \approx 157^\circ$; see Fig.~\ref{fig:costhetadft}. We observe the erosion of oscillations in the density profile and
the growth of a depleted region of density at each wall as $\epsilon_w$ is reduced in this range. This is accompanied by the
smoothing of oscillations in the local compressibility. As $\epsilon_w$ is reduced, the height of the maximum in the local
compressibility increases strongly and its location shifts to larger distances from the wall, following the position of
the maximum gradient of the density profile. These trends are consistent with DFT results in Ref.~\cite{Evans:2015aa} for a different
(SR) wall-fluid potential but pertaining to a similar range of contact angles. It is important to note that the results
in Fig.~\ref{fig:mariafig} for a single LR wall correspond to tiny values of the wall-fluid attraction where the drying film is very
thick; there we test the detailed predictions of the binding potential description in the limit of drying. In Fig.~\ref{fig:dftslitprofiles} we
are examining the overall changes of the density profiles and local compressibility as the substrate becomes more
solvophobic and the contact angle becomes very large. We should also note that the results in Fig.~\ref{fig:dftslitprofiles} are for the liquid
at bulk coexistence, i.e. $\delta\mu = 0^+$. For the confined fluid this state is metastable w.r.t. capillary evaporation. The
latter would occur at values of $\beta\delta\mu$ that are typically about $0.04$ --see Sec.~\ref{sec:pdslit}. Within DFT there is no difficulty in
probing these metastable `liquid' states which correspond to local minima of the excess grand potential. This is illustrated, for
smaller wall separations, in Fig. 8 of Ref.~\cite{Evans:2015aa}.

\section{Simulation methods} 
\label{sec:simmeth}
We employ GCMC simulation, which is well suited to studying fluids at vapor-liquid coexistence both in the bulk
\cite{Wilding1995} and in confinement \cite{Cracknell:1993aa,maciolek2003,Grzelak:2008ve,Rane:2011ly}. Within this
framework one prescribes the temperature $T$ and chemical potential $\mu$, while the particle number $N$ fluctuates. The
relevant observables are the probability function $P(\rho)$ of the total density $\rho=N/V$ and, for a confined system,
the density profile $\rho(z)$.

Although $\rho$ fluctuates in our simulations, close to coexistence state points sampling problems can arise due to the free
energy cost of traversing the mixed phase (interfacial) states that separate pure vapor from pure liquid. This cost is
manifest as a deep valley of low probability in $P(\rho)$ which, on simulation timescales, traps the sampling in one
phase. To overcome this problem we have implemented biasing techniques \cite{berg1992}, utilizing a weight function that
is calculated from the transition matrix \cite{Smith:1995kx}. The role of the weight function is to enhance the sampling
of mixed states, ie. to remove the sampling barrier. The effects of the biasing can subsequently be unfolded exactly
from distributions of observables.

Sufficiently close to the bulk vapor-liquid critical point, the requisite weights can be taken to be a function of the
total density $\rho$. However, at low temperature coexistence points (such as used in the present work) this
approach breaks down due to the appearance of finite-size induced first order phase transitions known as droplet
transitions \cite{Binder:2012ez}. For the purpose of biasing through these transitions, $\rho$ is not a good order
parameter and more effective alternatives must be sought. Suitable substitutes have recently been proposed by one of us
\cite{Wilding:2016aa}, and these were adopted in the present work.

GCMC is most efficient when deployed in conjunction with histogram extrapolation~\cite{Ferrenberg1989}. This permits the
results from a simulation performed at one set of model parameters, e.g. $T,\mu$ and wall strength $\epsilon_w$, to be
reweighted to provide estimates of observables at nearby parameters, without recourse to further simulation. In the
present work we have used histogram extrapolation to measure the local compressibility $\chi(z)$ from the $\mu$
dependence of the density profile $\rho(z)$. We have also used it in conjunction with results for $P(\rho)$ for a fully
periodic system to obtain accurate estimates of the coexistence chemical potential at a given temperature: tuning $\mu$
until the equal peak weight criterion is satisfied \cite{Wilding1995}.

\section{Results from simulations}
\label{sec:simresults}

\subsection{$P(\rho)$ and the contact angle}

\begin{figure}[h]
\includegraphics[type=pdf,ext=.pdf,read=.pdf,width=0.94\columnwidth,clip=true]{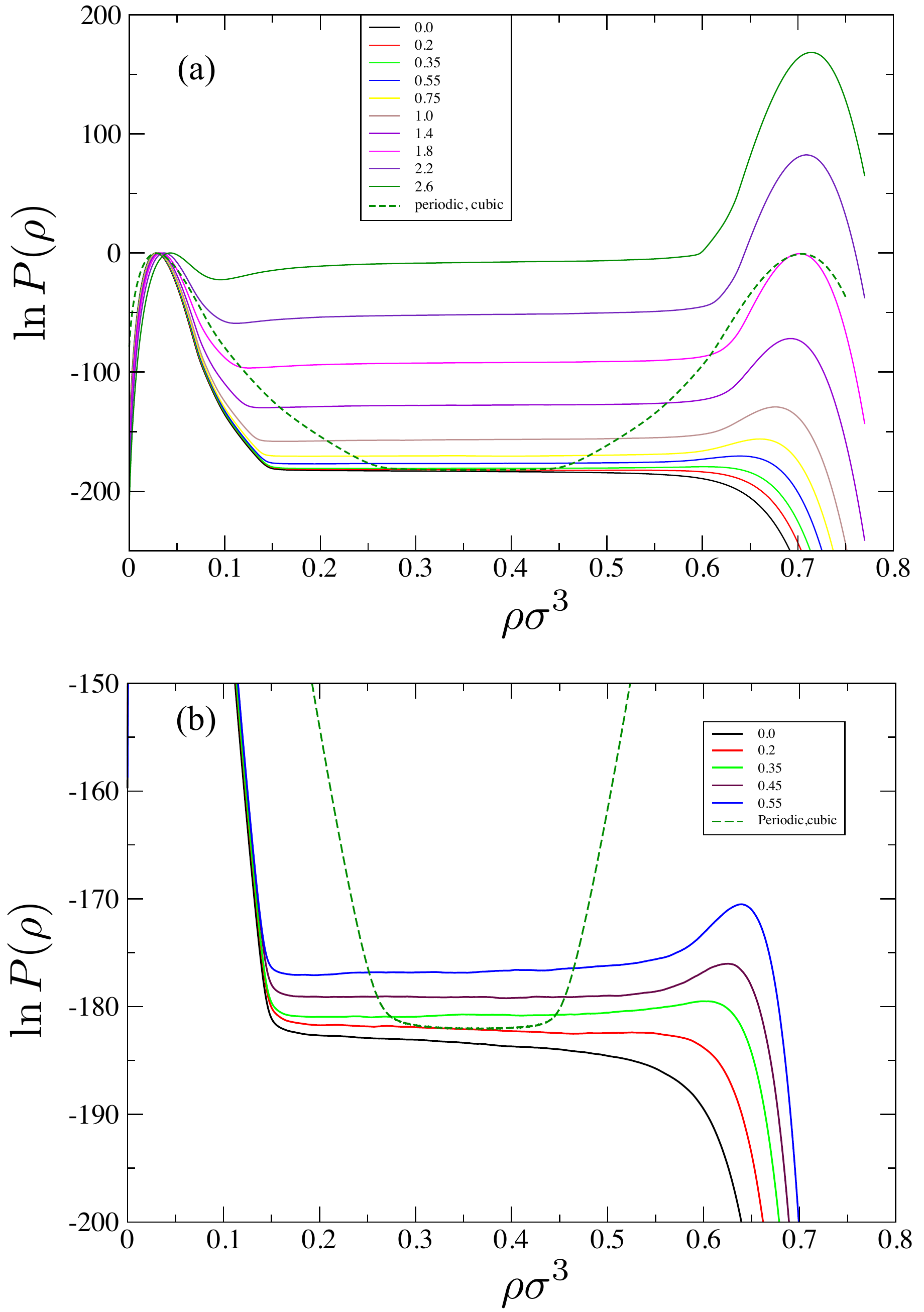}
\caption{ (a) The form of $\ln P(\rho)$ for the modified 9-3 potential at various $\epsilon_w$ as listed in the key.  The system size is $L=15\sigma, D=30\sigma$ and the temperature $T=0.775T_c$. (b) A close up of the region close to drying.   Also shown in both cases (dashed line) is $\ln P(\rho)$ measured for a fully periodic system of size $V=(15\sigma)^3$. }
\label{fig:LJ_dists}
\end{figure}

The contact angle as a function of wall strength $\epsilon_w$ can be obtained from the measured form of the density
probability function $P(\rho)$ in both a fully periodic system and in the slit. Since $P(\rho)$ can vary over many
decades it is convenient to work with its logarithm. Doing so has the additional advantage that the latter links
directly to the grand potential which is given by $\beta \Omega(\rho)=-\ln P(\rho)$.  Note that in all the simulation results that we present, 
$\epsilon_w$ refers to the wall-fluid potential strength measured in units of $k_BT$, i.e. for (\ref{eq:LRpot}) 
this quantity denotes $\epsilon_w\epsilon_{LJ}/k_BT=1.0877\epsilon_w$ and for
 (\ref{eq:SRpot}) denotes the well-depth in units of $k_BT$.  Figure~\ref{fig:LJ_dists}(a) shows
our GCMC results for $\ln P(\rho)$ for the slit system at vapor-liquid coexistence at a range of wall strengths
$\epsilon_w$ which span the regime from wetting to drying. The data shown is for the modified LR potential, but a
similar scenario plays out for the SR potential. For sufficiently large $\epsilon_w$, a double peaked structure is
evident in $P(\rho)$. The low density peak (of height $P_{\rm vap}$) corresponds to the system in a capillary vapor
phase, while the high density peak (of height $P_{\rm liq}$) corresponds to the capillary liquid phase. For large
$\epsilon_w$ the liquid has the higher peak (i.e. is the stable phase), but as $\epsilon_w$ is reduced, the height of
the liquid peak diminishes progressively, until it becomes metastable with respect to the vapor. The liquid peak height
continues to diminish as $\epsilon_w$ is decreased until eventually it disappears into a plateau. At still smaller
$\epsilon_w$, $P(\rho)$ is a monotonically decreasing function in the region of liquid-like densities, as shown in
Fig.~\ref{fig:LJ_dists}(b). Also included in Fig.~\ref{fig:LJ_dists} (dashed line) is the coexistence form of $P(\rho)$
for a fully periodic cubic system of side $L$ that matches the linear dimension of the planar walls in the slit system,
c.f. eq.~(\ref{eq:slitvol}). For this latter system, $P(\rho)$ exhibits a pair of equal peaks of probability $P_{\rm
max}$, corresponding to the respective pure phase states. These are separated by a central plateau (of probability
$P_{\rm min}$) corresponding to mixed phase (interfacial) states.

As is well established, the ratio of peak to valley probabilities in the fully periodic case provides an accurate estimate of the vapor-liquid surface tension $\gamma_{lv}$ \cite{Binder:1982ck,Errington2003}:

\be 
\gamma_{vl}=(2\beta L^2)^{-1}\ln(P_{\rm max}/P_{\rm min})\:,  
\label{eq:gamlv}
\ee 
where $\beta=(k_BT)^{-1}$. Similarly the ratio of peak heights for the distributions in the slit system provides a measure of the surface tension difference \cite{Muller:2000fv}:

\be
\gamma_{wv}-\gamma_{wl}=-(2\beta L^2)^{-1}\ln(P_{\rm vap}/P_{\rm liq})\:.
\label{eq:gamdiff}
\ee
Accordingly, one can simply read off these quantities directly from the measured forms of $P(\rho)$ and insert them into Young's equation (\ref{eq:Young}) in order to obtain an estimate of the contact angle as

\be
\cos(\theta) = \frac{ \gamma_{wv}-\gamma_{wl}}{\gamma_{vl}}\:.
\label{eq:Young1}
\ee 
\ 

\begin{figure}[h]
\includegraphics[type=pdf,ext=.pdf,read=.pdf,width=0.94\columnwidth,clip=true]{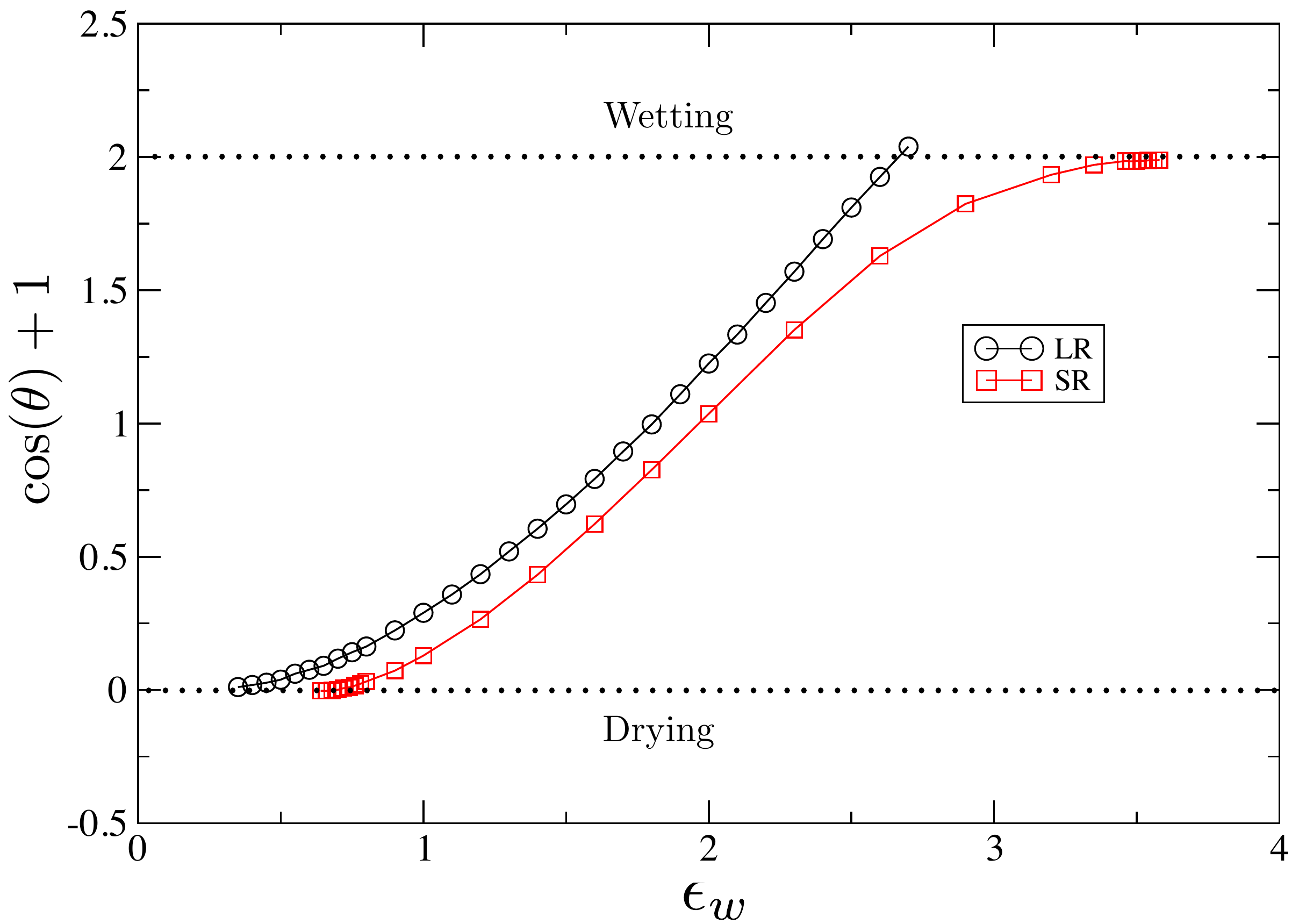}
\caption{ GCMC results for $\cos(\theta)+1$ versus
    $\epsilon_w$ for the SR and LR wall potential at vapor-liquid coexistence for $T=0.775T_c$. The system size is
    $L=15\sigma, D=30\sigma$. Lines are guides to the eye. }

\label{fig:costhetasim}
\end{figure}

This methodology can be used to estimate $\cos(\theta)$ as a function of wall strength $\epsilon_w$ for both the LR and
SR wall-fluid potentials. The results are shown in Fig.~\ref{fig:costhetasim} and should be compared with those of the DFT
calculations of Fig.~\ref{fig:costhetadft}. A number of pertinent features are apparent. First we note that for both the
LR and SR cases, $\cos(\theta)$ appears to approach $-1$ tangentially both in simulation and DFT. This behaviour is
expected for a {\em critical} drying transition \cite{Dietrich:1988et}, see Secs.~\ref{sec:phenomenology} and
\ref{sec:bp}. Second, in both the LR and SR cases, the simulations appear to indicate that drying occurs for a small but
non-zero wall strength $\epsilon_w$, the value of which is substantially smaller for the LR case than the SR case. This
appears to signal a qualitative discrepancy with DFT which unambiguously predicts (in accord with our binding potential
calculations-see Sec.~\ref{sec:bp}) that for the LR case critical drying occurs at $\epsilon_w=0$. We shall return to
this discrepancy. Third, there are clear qualitative differences between drying and wetting for the LR case: While for
the SR case $\cos(\theta)$ approaches the wetting limit $\cos(\theta)=1$ tangentially, similar to drying, and therefore
indicating critical wetting, for the LR case the approach to this limit is with non-zero gradient, indicative of first
order wetting \cite{Dietrich:1988et}. The DFT results shown in Fig.~\ref{fig:costhetadft} are consistent with this
finding.

Beyond their utility for determining contact angles, the relations (\ref{eq:gamlv})-(\ref{eq:Young1}) permit one to forge
the link between principal features of Fig.~\ref{fig:LJ_dists} and the surface phase diagram (cf.
Fig.~\ref{fig:schematicpd}). The wetting point ($\cos(\theta)=1$) occurs for $P_{\rm liq}/P_{\rm vap}=P_{\rm max}/P_{\rm
min}$, which in the LR case occurs for $\epsilon_w\approx 2.6$. From fig.~\ref{fig:LJ_dists}(a) one sees that while the
liquid peak is strongly stable at this point, there is a {\em metastable} vapor peak which corresponds to a local free
energy minimum. This minimum serves to bind the vapor phase to the wall at the wetting transition and hence --and in
accord with the observed behaviour of the contact angle-- the wetting transition is first order in this
system~\footnote{A weak bulge is evident in $\ln P(\rho)$ for $\rho\sigma^3\simeq 0.2$ at the wetting point; we
speculate that this may be a signature of the incipient prewetting transition.}.

The partial wetting regime, defined by $0<\cos(\theta)<1$, occurs for $1.8\lesssim\epsilon_w\lesssim2.6$. Its lower
boundary is marked by the `neutral' wall for which $P_{\rm vap}=P_{\rm liq}$. This heralds entry into the partial drying
regime ($-1\leq\cos(\theta)<0$) within which, for confinement within a slit, the liquid phase is metastable with respect
to the vapor. Drying occurs for $\cos(\theta)=-1$ and corresponds to the wall strength for which $P_{\rm liq}/P_{\rm
vap}= P_{\rm min}/P_{\rm max}$. Fig.~\ref{fig:LJ_dists}(b) shows that the value of $\epsilon_w$ at which this equality
is satisfied coincides closely with the point at which the liquid peak disappears smoothly into a plateau. We can
therefore {\em provisionally} identify the drying point as that wall strength for which the liquid peak vanishes. The
smoothness with which this occurs is consistent with the arguments presented above advocating that drying is critical.

\subsection{Finite-size scaling: pinning down the drying transition}
\label{sec:fse}

\begin{figure}[h]
\includegraphics[type=pdf,ext=.pdf,read=.pdf,width=0.94\columnwidth,clip=true]{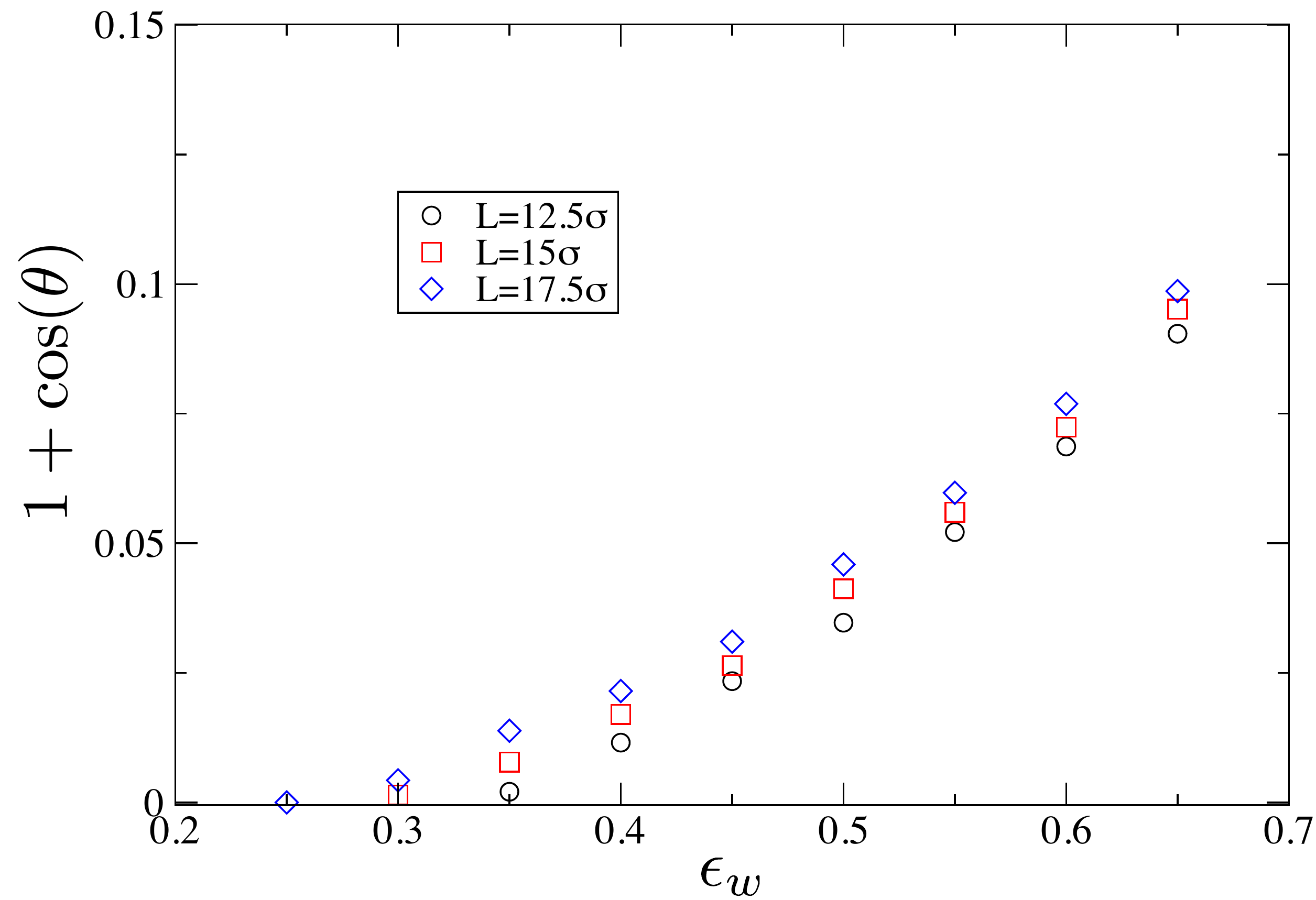}
\caption{ GCMC results for $\cos(\theta)+1$ as a function of the LR wall strength $\epsilon_w$ for system sizes $D=30\sigma, L=12.5\sigma, 15\sigma, 17.5\sigma$. } 
\label{fig:costhetaLdep}
\end{figure}

Contact angle measurements provide an accurate indication of {\em the order} of a surface phase transition. However in
the case of a critical surface phase transition, they do not yield accurate estimates for its location, nor for the
associated critical exponents. The problem goes beyond the inherent difficulty of estimating the wall strength for which
$\cos(\theta)=\pm 1$ when the approach to this limit is tangential; the main difficulty is one of finite-size effects.
This is demonstrated in Fig.~\ref{fig:costhetaLdep} which shows our GCMC estimates of $\cos(\theta)$ for values of
$\epsilon_w$ in the vicinity of the critical drying transition, for three values of $L$; recall $L^2$ is the wall area.
The data clearly show that the apparent drying point $\cos(\theta)=-1$ shifts systematically to lower values of
$\epsilon_w$ as $L$ increases. This observation is important and we return to it later.

To clarify the nature of the near-critical finite-size effects, it is useful to return to the density distribution
$P(\rho)$. In examining this quantity we shall exclude the low density region where the vapor peak occurs. This peak corresponds to the capillary
evaporation transition that occurs when two vapor-liquid interfaces unbind from the wall and wander to the
slit centre where they annihilate. By excluding it from the sampling we can focus on the behaviour at higher
(liquid-like) densities, which are the ones relevant for critical drying. A further advantage is that we can switch
from a logarithmic to a linear scale which is more revealing as regards exposing the character of the
criticality.

\begin{figure}[h]
\includegraphics[type=pdf,ext=.pdf,read=.pdf,width=0.94\columnwidth,clip=true]{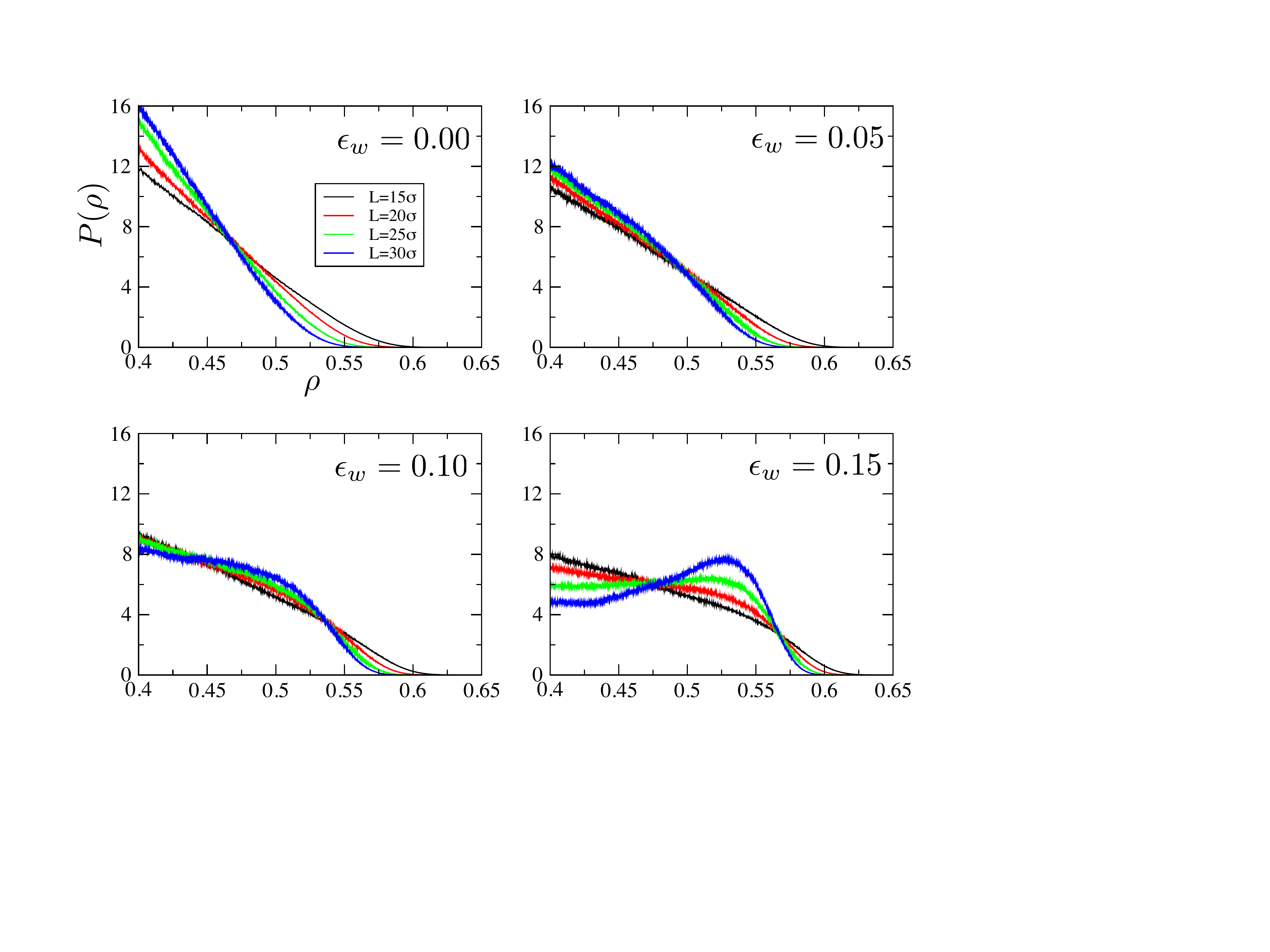}
\caption{GCMC results for $P(\rho)$ for the LR wall
    potential for $D=30\sigma$ and various $L$ at a selection of
    near-critical drying values of $\epsilon_w$.  }
 
\label{fig:LRFS}
\end{figure}
 
Fig.~\ref{fig:LRFS} shows the measured forms of $P(\rho)$ for the system with LR wall-fluid interactions at a selection
of system sizes $L$ and wall strengths $\epsilon_w$. One sees that for sufficiently large $\epsilon_w$ and $L$,
$P(\rho)$ exhibits a liquid peak. On reducing $\epsilon_w$, this peak disappears into a plateau. On further reduction
$P(\rho)$ becomes monotonically decreasing with a bulge which gradually diminishes until, at $\epsilon_w=0$, the
distribution comprises a linear part and a tail. The interesting feature of Fig.~\ref{fig:LRFS} is that the presence or
otherwise of a liquid peak at a given $\epsilon_w$ depends on the value of $L$. Even when $\epsilon_w$ is sufficiently
small that only a bulge (but no peak) is seen, the bulge grows over the range of accessible $L$, suggesting that a peak
would form were sufficiently large values of $L$ attainable.

The range of $\epsilon_w$ over which the distribution evolves from having a peak to becoming linear with a tail,
decreases with increasing $L$ indicating scaling behaviour. Only for $\epsilon_w=0$ is the form of $P(\rho)$ scale
invariant, ie. no peak begins to form as $L$ is increased. In view of this behaviour we believe that $\epsilon_w=0$
marks the true critical drying point for the system with LR wall-fluid interaction and that in the thermodynamic limit a
liquid peak (indicating partial drying) will occur for all $\epsilon_w>0$. The finite-size dependence of the value of
$\epsilon_w$ for which a peak first occurs feeds through to the observed finite-size shift in the apparent drying point
as measured via the contact angle calculation (fig.~\ref{fig:costhetaLdep}). Recall that both DFT
(figs.~\ref{fig:mariafig}-\ref{fig:costhetadft}) and binding potential calculations (Sec.~\ref{sec:bp}) also predict
critical drying for $\epsilon_w=0$. Note also that for $\epsilon_w=0$, $W_{LR}(z)$ in (\ref{eq:LRpot}) reduces to the
hard wall potential which we know leads to complete drying \cite{Henderson:1985aa,Oettel:2005aa}.

\begin{figure}[h]
\includegraphics[type=pdf,ext=.pdf,read=.pdf,width=0.6\columnwidth,clip=true]{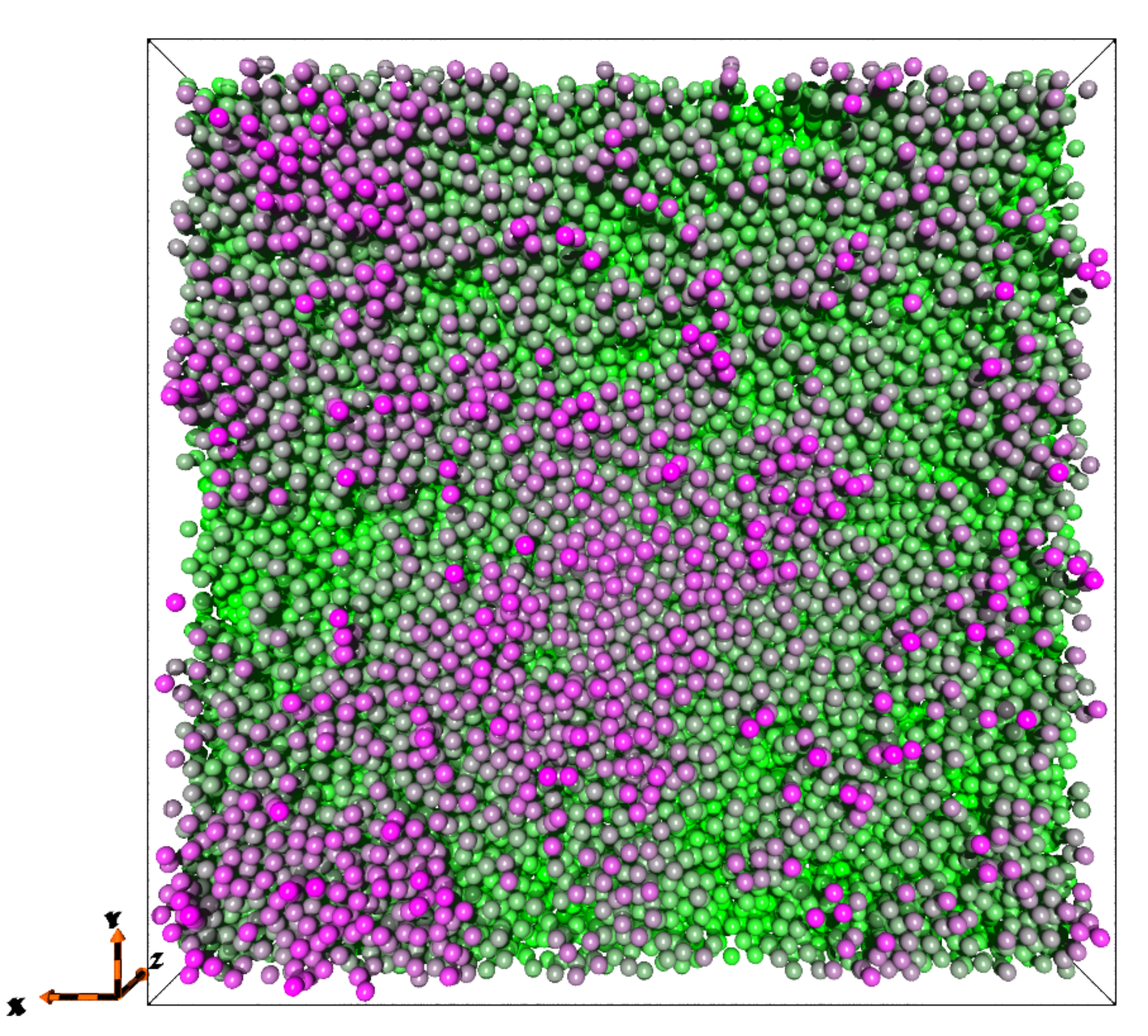}
\caption{Simulation snapshot for a system with
    $L=50\sigma, \epsilon_w=0.2$. Particles are color coded
    according to their distance from the wall at $z=0$, with purple closest to the wall and green furthest away. A large
    correlation length is manifest in the vapor close to the wall.}
\label{fig:dryconf}
\end{figure}

The form of $P(\rho)$ at criticality, namely a linear part with a tail, is essentially a universal finite-size scaling
(FSS) function, albeit a trivial one. The simplicity of its form stands in stark contrast to critical point FSS
functions for the order parameter found at bulk critical points (see eg. \cite{Wilding1995}). This difference highlights
what we believe to be a fundamental difference between bulk and surface criticality in 3d as probed by simulations,
namely that one can approach, but never quite reach a surface critical point.

To elaborate, consider the behaviour approaching the critical drying point $\epsilon_w\to \epsilon_{wd}^+$. As criticality is neared, the parallel
correlation length grows like $\xi_\parallel\sim (\epsilon_w-\epsilon_{wd})^{-\nu_{\parallel}}$ (see
eq.~\ref{eq:xipardiv} and Sec.~\ref{sec:profiles} below). Physically, one can regard $\xi_\parallel$ as
reflecting the lateral size of `bubbles' of the incipient (vapor) phase that form at the wall. The divergence
in the bubble size as $\epsilon_w\to \epsilon_{wd}^+$ implies that $\xi_\parallel$ can grow up to the system size
$L$. However, the situation is different for the bubble thickness in the perpendicular direction. This
thickness is given by $\xi_\perp$, the surface roughness which also, in principle, diverges (albeit
logarithmically) as $\epsilon_w\to \epsilon_{wd}^+$. However, unlike $\xi_\parallel$, the broken translational
symmetry perpendicular to the walls implies that in 3d simulations $\xi_{\perp}$ is strongly dampened by
finite-size effects. General capillary wave arguments e.g. \cite{Gelfand:1990ve,Dietrich:1988et,Evans:1992jo}
for a single unbinding vapor-liquid interface predict that the surface roughness $\xi_\perp\simeq
\sqrt{(k_BT/2\pi\gamma_{lv})\ln (L/\xi_b)}$. Thus the surface roughness depends on the finite {\em lateral}
dimension of the system and owing to the strong $\sqrt{\ln L}$ dampening one expects that  for
currently accessible system sizes the bubble thickness does not become large on the scale of the
particle diameter (or indeed the bulk correlation length $\xi_b$). A configurational snapshot of the emerging vapor-liquid
interface at the wall (Fig.~\ref{fig:dryconf}) in which particles are colored according to their distance from
the wall,  confirms qualitatively this picture. Observing the correlated regions of purple shaded particles lying close to the wall and the green shaded particles further from the wall, 
we note that there is a large but finite $\xi_\parallel$ manifest in the large fractal bubbles of `vapor'
which almost span the system in the lateral dimension. However, the perpendicular extent of these bubbles is
microscopic, extending only a few particle diameters away from the wall. We discuss the accompanying density profile in the next subsection.

As $\xi_\parallel\to L$ for some $\epsilon_w>\epsilon_{wd}$, a vapor bubble spans the wall allowing the liquid to
unbind and form a free `slab', surrounded by vapor. In terms of the form of $P(\rho)$, this happens when the
liquid peak --which corresponds to a local free energy minimum that serves to binds the liquid to the wall--
disappears. Recall, however, that the vanishing of the liquid peak occurs at the same value of $\epsilon_w$ for
which the contact angle measurements predict $\cos(\theta)=-1$. Thus the unbinding process can be viewed as
{\em premature} drying induced by the finite system size. We shall denote the wall strength for which it
occurs by $\epsilon_{wd}(L)$. Since the unbound liquid slab can fluctuate freely away from the wall, premature drying marks a spontaneous
{\em loss} of the near-critical state at $\epsilon_w\approx\epsilon_{wd}(L)$. This state of affairs contrasts
starkly with the situation for simulations of bulk critical phenomena where correlations diverge isotropically
and critical fluctuations can be sampled right up to criticality. It appears not to have been recognized previously.

The existence of the non-critical state allows one to
rationalize the form of $P(\rho)$ at critical drying. Owing to the dampening of capillary fluctuations, the
surface of the detached liquid slab is rather sharp and localized and hence the slab thickness (in the $z$-direction)
is proportional to $\rho$. Accordingly, the linear decrease of $P(\rho|\epsilon_w=0)$ seen at low to moderate
densities in Fig.~\ref{fig:LRFS} arises simply from the `entropic repulsion' of the slab and the wall: the
number of positions for the slab center along the $z$ axis that are allowed by the presence of the wall,
varies linearly with slab thickness. The high density tail of $P(\rho)$ on the other hand reflects the free
energy cost of pushing the liquid up against the wall, the act of which quenches the parallel density
fluctuations. Its $L$ dependence arises --as shown in fig S2. of \cite{Evans:2016aa}-- from a constant repulsive pressure
on the liquid-vapor interface by the wall, giving rise to a force which scales simply with the wall area
$L^2$. 

\begin{figure}[h]
\includegraphics[type=pdf,ext=.pdf,read=.pdf,width=0.94\columnwidth,clip=true]{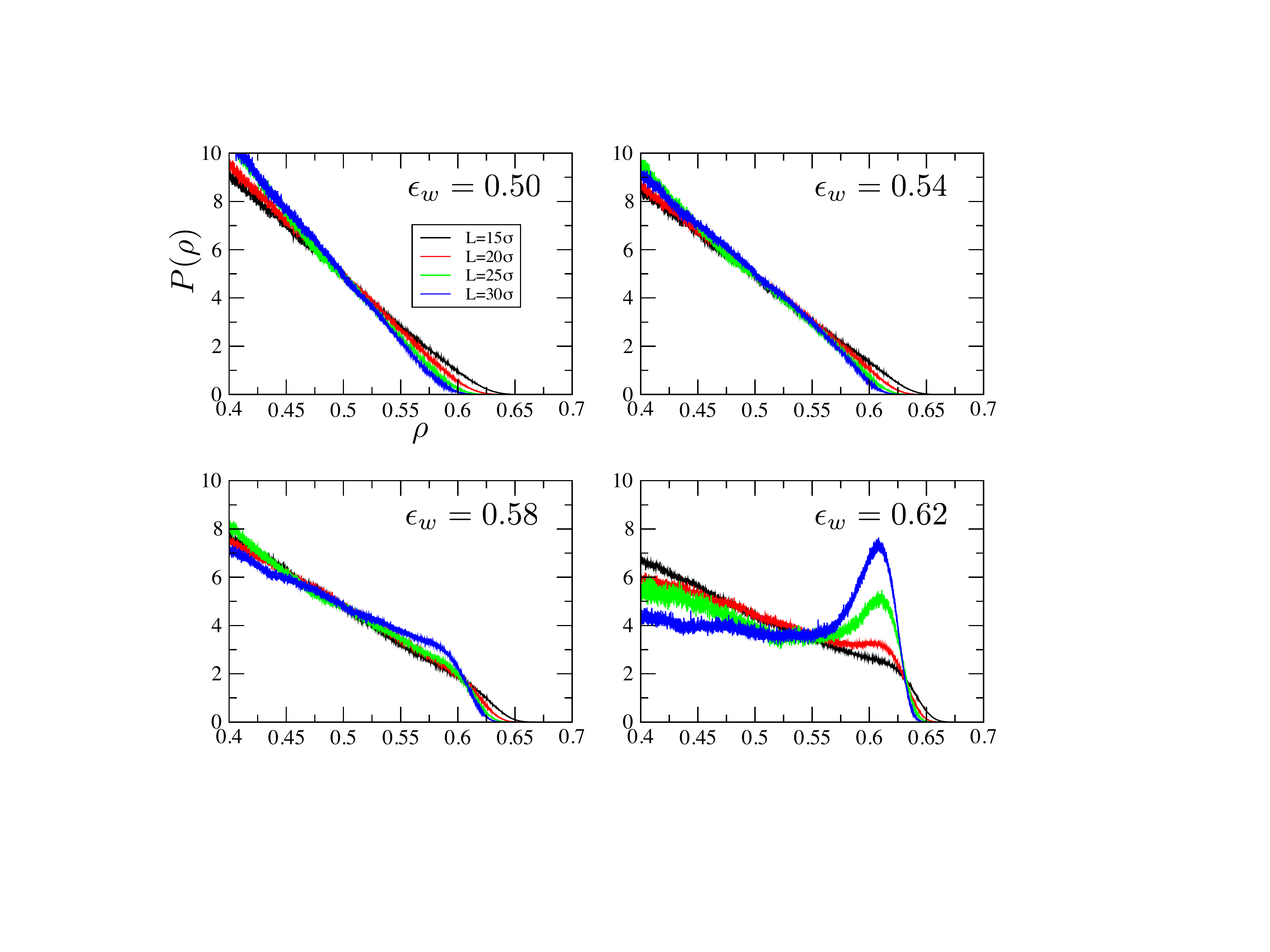}
\caption{GCMC results for $P(\rho)$ for the SR wall
    potential for $D=30\sigma$ and various $L$ at a selection of
    near-critical drying values of $\epsilon_w$. }
 
\label{fig:SWFS}
\end{figure}

The critical wall strength, $\epsilon_{wd}$, is determined most accurately as the largest
value of $\epsilon_w$ for which $P(\rho)$ assumes an $L$-independent form. This value can differ substantially from $\epsilon_{wd}(L)$. For the LR system we find $\epsilon_{wd}=0$, in agreement with the prediction of binding potential and DFT calculations. However, for a system having SR wall-fluid interactions (as described in Sec.~\ref{sec:modelmeth}),  binding potential calculations, as well as DFT and GCMC estimates of contact angles  (cf. Eq.~ (\ref{eq:SRcostheta}) and Figs.~\ref{fig:costhetadft},\ref{fig:costhetasim}) predict that a critical drying transition occurs for a non-zero attractive wall strength. The value of the critical drying point is not known {\em a-priori} from a binding potential analysis in this case and hence it is interesting to see whether an accurate determination can be made by examining the $\epsilon_w$ and $L$ dependence of $P(\rho)$.  Fig.~\ref{fig:SWFS} shows that the behaviour mirrors qualitatively that found in the LR case (Fig.~\ref{fig:LRFS}). The main difference is that on decreasing $\epsilon_w$ from large values, a {\em non-zero} value of $\epsilon_w$ is reached below which $P(\rho)$ exhibits a linear part and a tail for all $L$. On the basis of the arguments given above we take this value to be the critical drying point which we estimate to occur for $\epsilon_{wd}=0.52(2)$. 
 
\subsection{Density and compressibility profiles and a Maxwell relation.} 
 ~\label{sec:profiles} 

The existence of premature drying identified above implies that the critical limit can be accessed only in the
thermodynamic limit $L\to\infty$. This begs the practical question: How closely can one approach criticality for a given
$L$ such that estimates of observables are representative of the thermodynamic limit in the near-critical region rather than the non-critical state?

\begin{figure}[h]
\includegraphics[type=pdf,ext=.pdf,read=.pdf,width=0.94\columnwidth,clip=true]{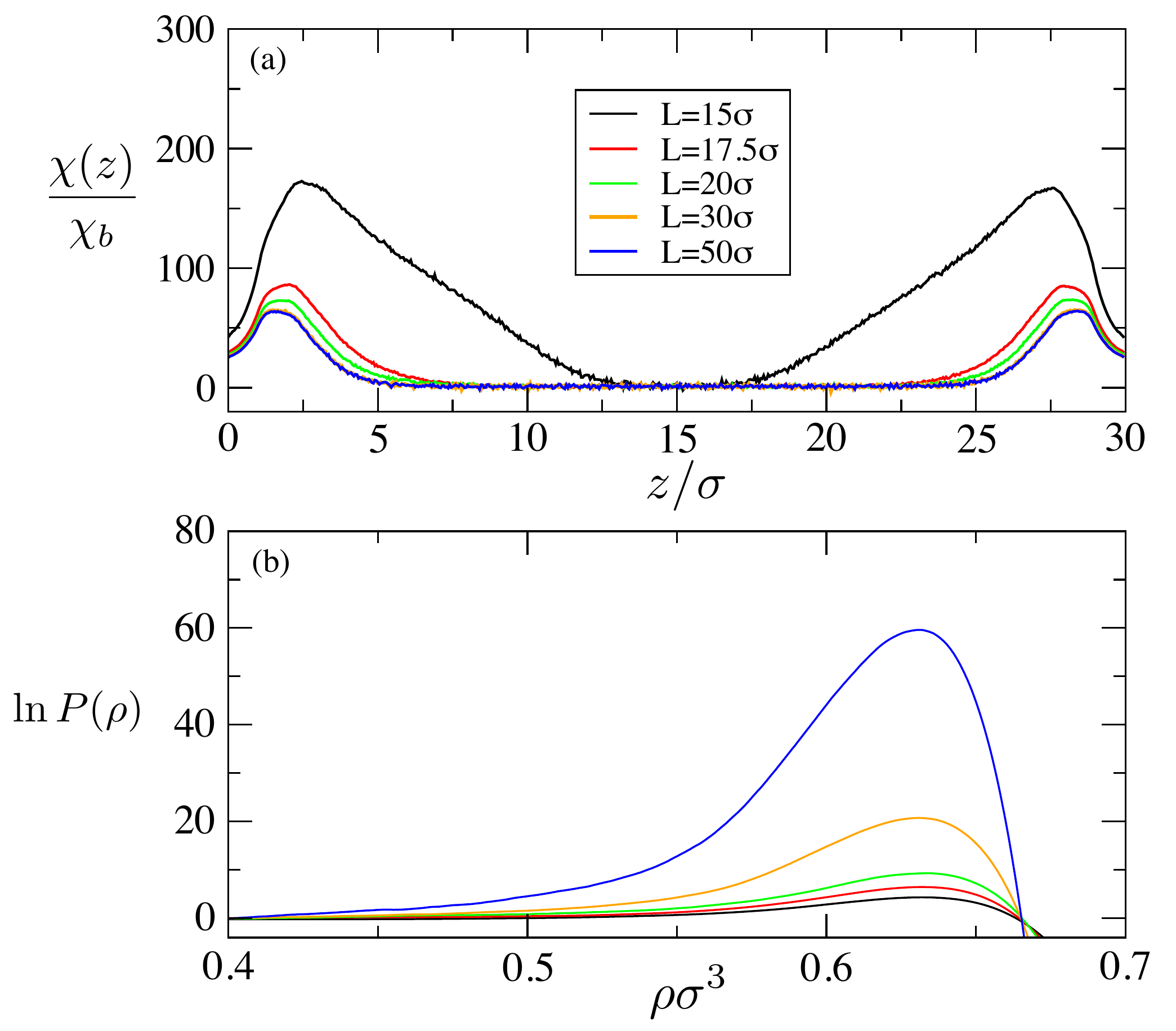}
\caption{{\bf (a)} GCMC results for the normalised compressibility profile $\chi(z)/\chi_b$ for the LR wall-fluid potential with $\epsilon_w=0.5$ for various system sizes $L$. {\bf (b)} The corresponding forms of $\ln P(\rho)$ in the density range corresponding to the metastable liquid peak. Note that the curves have been shifted vertically so that they coincide at $\rho\sigma^3=0.4$ in the plateau region.}
\label{fig:comp_finitesize}
\end{figure}

To answer this question we have investigated the near-critical behaviour of the local compressibility and density profiles.
Fig.~\ref{fig:comp_finitesize}(a) compares the form of $\chi(z)$ for a selection of values of $L$ for the LR system at
the near-critical wall strength $\epsilon_w=0.5$.  A dramatic finite-size dependence is apparent. Specifically, as $L$ is
increased from small values, $\chi(z)$ decreases strongly, before converging for sufficiently large $L$. Such a decrease
is at first sight most surprising because in bulk systems the total compressibility generally increases with system size in
the vicinity of a critical point. The origin can be traced to a smaller free energy cost for fluctuations to lower
densities compared to those to higher densities, as manifest in the finite-size forms of $\ln P(\rho)$ (Fig.~\ref{fig:comp_finitesize}(b)). In the partial drying regime, the system occupies states whose densities lie under
the liquid peak shown in this figure. However, the shape of the peak is strongly {\em asymmetric}, with a tail extending
to lower densities which runs smoothly into the plateau associated with the non-critical fluctuations of an unbound
liquid slab. The tail reflects the relative `softness' of fluctuations that reduce the density near the wall.
Fig.~\ref{fig:comp_finitesize}(b) also shows that the liquid peak height (as measured from the plateau) grows with
increasing $L$~\footnote{For a given $\epsilon_w$, the height of the liquid peak measured from the plateau corresponds to
the free energy cost per unit wall area of forming a vapor-liquid interface by removing the liquid from the wall, ie. it
is proportional to $\gamma_{wv}-\gamma_{wl}+\gamma_{vl}$. }. Accordingly, the extent to which the density fluctuations
can escape the top of the liquid peak and sample the tail and plateau region decreases with increasing $L$. It follows
that for small $L$ the sampling includes contributions from slab fluctuations (and/or their precursors in the tail
region) resulting in a spurious enhancement of the compressibility. 

Given this insight, it is interesting to reassess the role of finite-size effects in previous simulation studies for
fluids near weakly attractive substrates. In Ref.~\cite{Evans:2015wo}, $\chi(z)$ was measured for SPC/E water as a
function of the attractive wall strength $\epsilon_w$. For small $\epsilon_w$, the form of $\chi(z)$ that was observed
is similar to that shown for $L=15\sigma$ in Fig.~\ref{fig:comp_finitesize}(a). In view of the smaller values of $L$
that were attainable for the water model compared to the current LJ system, it is now clear that this compressibility
profile was affected by finite-size effects. Remaining with SPC/E water, we note reports of asymmetry in the form of the
probability function of the fluctuating density within a subvolume located close to a large hydrophobic solute particle
\cite{Patel:2010dz,Patel:2012aa}. The findings shown in Fig.~\ref{fig:comp_finitesize}(b) help rationalize this
observation in terms of finite-size effects.

\begin{figure}[h]
\includegraphics[type=pdf,ext=.pdf,read=.pdf,width=0.94\columnwidth,clip=true]{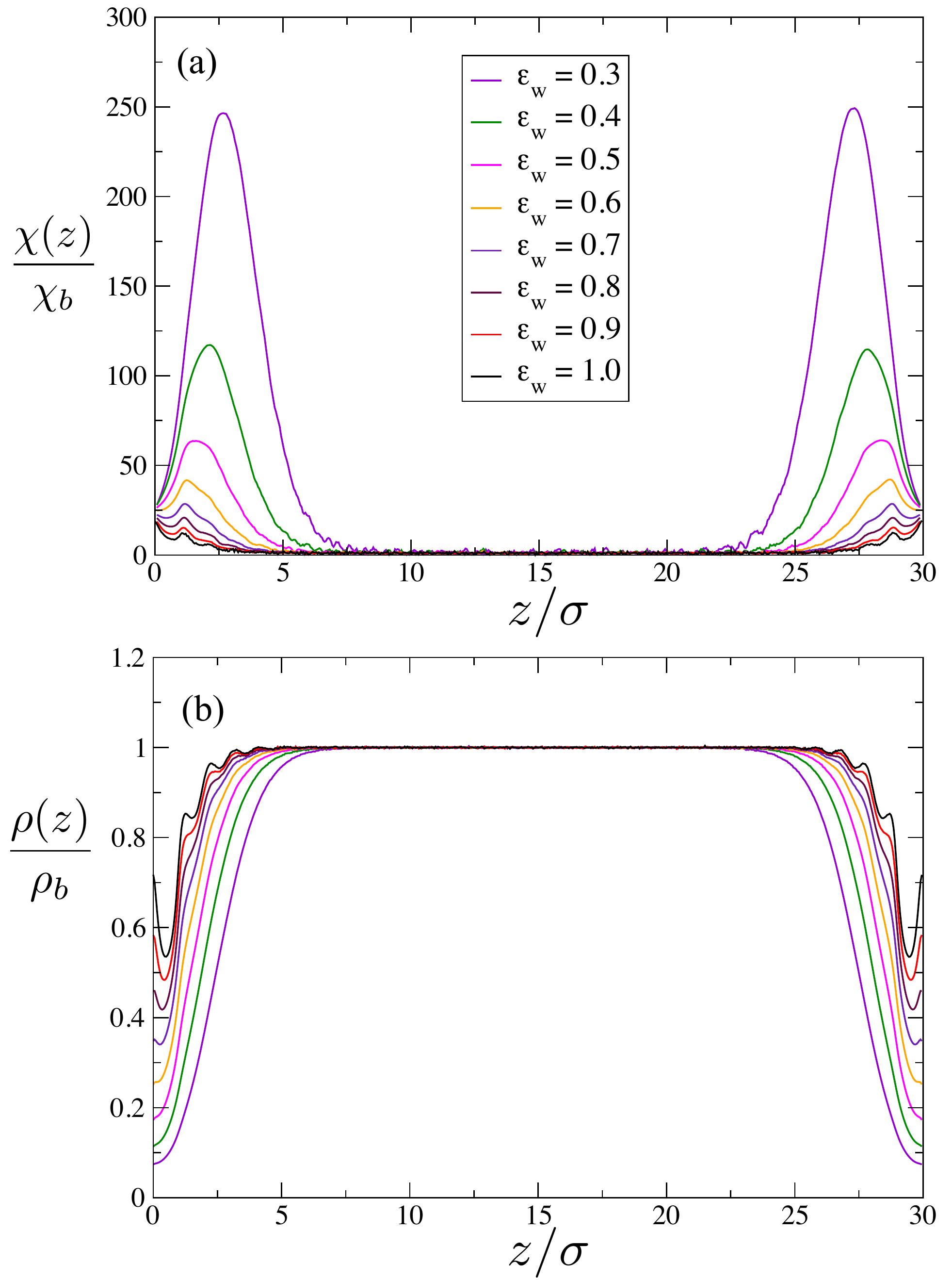}
\caption{{\bf (a)} GCMC results for the normalised compressibility profile $\chi(z)/\chi_b$  at vapor liquid coexistence for various strengths $\epsilon_w$ of the LR system as given in the key. The system size is $L=50\sigma,D=30\sigma$, and the temperature is $0.775T_c$.  {\bf (b)}  Corresponding results for the normalised density profile $\rho(z)/\rho_b$.}
\label{fig:profiles}
\end{figure}

The upshot of our analysis of finite-size effects is that in order to obtain estimates of observables 
that are representative of the thermodynamic limit, one must ensure that $L$ is sufficiently large for the prescribed $\epsilon_w$ that
the liquid peak is high. This in turn requires $\xi_\parallel\ll L$. Measurements of $\rho(z)$ and
$\chi(z)$ that have been found to be $L$-independent are shown in Fig.~\ref{fig:profiles} for $L=50\sigma$. For
this rather large value of $L$, finite-size effects are found to be small provided $\epsilon_w\gtrsim 0.3$ (recall that
criticality is at $\epsilon_w=0$ for the LR system). Within this region we observe clear evidence of strong
near-critical fluctuations: the local compressibility near the wall exceeds its bulk value by a factor in
excess of $200$, cf. Fig.~\ref{fig:profiles}(a).  Additionally, a growing drying layer is associated with the density profile
of Fig.~\ref{fig:profiles}(b). Note, however, that owing to its very weak critical divergence -- see (\ref{eq:leq}), the drying layer
thickness does not attain more than a few particle diameters, for these values of $L$, even when $\xi_\parallel$ is an order of magnitude larger. 

Comparison with the DFT results in Fig.~\ref{fig:dftslitprofiles} are revealing. We note first that in the simulations
the density and $\chi(z)$ profiles very close to the walls have different shapes from those obtained in DFT. The latter
decrease smoothly to zero as $z\to 0$, reflecting the soft $wf$ repulsion in (\ref{eq:LRpot}). Since the $wf$ potential
employed in simulations is the modified form with its minimum at the hard-wall, $z=0$, the density profiles, for larger
values of $\epsilon_w$, are increasing as $z\to 0$. However, this difference is not important for small values of
$\epsilon_w$ when the drying layer has developed. The density profiles then have very similar forms and integrated
quantities such as the adsorption should not depend on the fine details of the $wf$ potential. $\chi(z$) goes to a
non-zero value at $z=0$ in simulation but the overall variation in shape with $\epsilon_w$ is similar to that in DFT.
What is striking is that for a similar thickness of drying layer the simulation results yield much larger maxima in
$\chi(z$).

\begin{figure}[h]
\includegraphics[type=pdf,ext=.pdf,read=.pdf,width=0.94\columnwidth,clip=true]{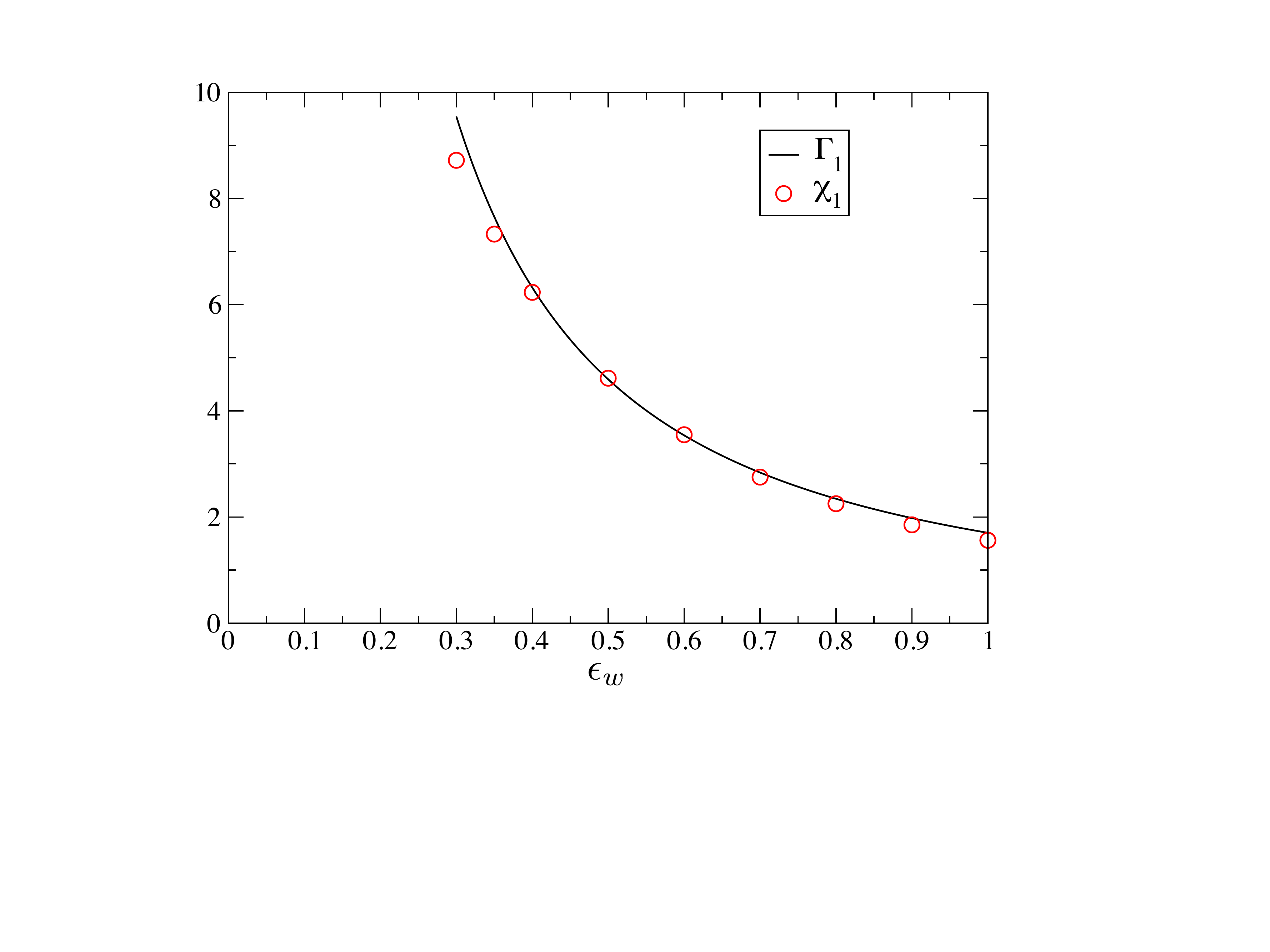}
\caption{A comparison of $\beta^{-1}\Gamma_1\sigma^2$ and $\beta^{-1}\chi_1\sigma^2$ for the LR system as related by the Maxwell relation eq.~(\ref{eq:maxwell}) with $L=50\sigma, D=30\sigma$. The temperature is $T=0.775T_c$. The line is a fit to $\beta^{-1}\Gamma_1$. Statistical errors are comparable with the symbol sizes.}
\label{fig:maxwell}
\end{figure}

As criticality is approached, finite-size effects begin to manifest themselves when $\xi_\parallel\simeq L$. We find
that the form of $\chi(z)$ is considerably more sensitive to changes in $L$ in this regime than is $\rho(z)$ -- an
effect that is traceable to the much stronger critical divergence of $\xi_\parallel$ compared to that of the drying film
thickness (or the adsorption $\Gamma$), cf. Eqs.~(\ref{eq:chizS}),(\ref{eq:leq}),(\ref{eq:logchi}). To probe further the relationship between
the two profiles, we have examined within simulation the Maxwell relation eq.~(\ref{eq:maxwell}), that links
$\Gamma_1\equiv\partial\Gamma/\partial \epsilon_w$ to the weighted compressibility $\chi_1\equiv\partial
\Theta/\partial\mu$. The adsorption $\Gamma(\epsilon_w)$ was obtained from $\rho(z)$ using (\ref{eq:Gammadef}) allowing
access to its numerical derivative $\Gamma_1$ in (\ref{eq:maxwell}). A fit to the latter is shown in
Fig.~\ref{fig:maxwell}, where it is compared with our measurements of $\chi_1$. One observes that at large $\epsilon_w$,
far from criticality, there are small discrepancies between the two quantities. These we attribute to the fact that when the
drying layer is thin $\rho(z)$ contains more structure and numerical integration is less accurate. On moving to smaller
$\epsilon_w$, for which $\xi_\parallel$ is large (but still small compared to $L$), there is excellent agreement between
$\Gamma_1$ and $\chi_1$ -- a finding that verifies our numerics. However, as the wall strength $\epsilon_w\approx 0.3$
is reached, a significant discrepancy starts to appear. Given the differing sensitivities of $\chi(z)$ and $\rho(z)$ to
finite-size effects noted above, we speculate that this discrepancy serves as an indicator that the limit $\xi\approx L$
has been reached and finite-size effects are significant. The extent to which the Maxwell relation holds therefore
appears to serve as a useful tool for diagnosing when finite size effects are significant.

 \subsection{Estimates of $\nu_{\parallel}$}
\label{sec:nu}

In Sec~\ref{sec:fse} the form of $P(\rho)$ was examined for state points near criticality. Precisely at criticality $P(\rho)$ comprises a linear part and a tail. The linear part occurs at low to moderate densities and arises from the entropic repulsion of the wall to the unbound liquid slab. The high density tail arises from the free energy cost of pushing the slab up against the wall. Neither of these phenomena is directly associated with
criticality, and thus one cannot expect $P(\rho)$ to exhibit non-trivial finite-size scaling (FSS) behavior as a whole. Rather, the signature of near critical fluctuations is manifest in the density range where the liquid is still (weakly) bound to the wall but exhibits strong parallel density fluctuations. This correspond to the liquid peak in Fig.~\ref{fig:LRFS}, the height of which depends on $\xi_\parallel$ and vanishes when $\xi_\parallel\approx L$ allowing the liquid slab to
unbind from the wall. Simple FSS dictates that this vanishing occurs not at $\epsilon_{wd}$ but at the larger effective value $\epsilon_{wd} (L) = \epsilon_{wd} + aL^{-1/\nu_\parallel}$. 

\begin{figure}[h]
\includegraphics[type=pdf,ext=.pdf,read=.pdf,width=0.94\columnwidth,clip=true]{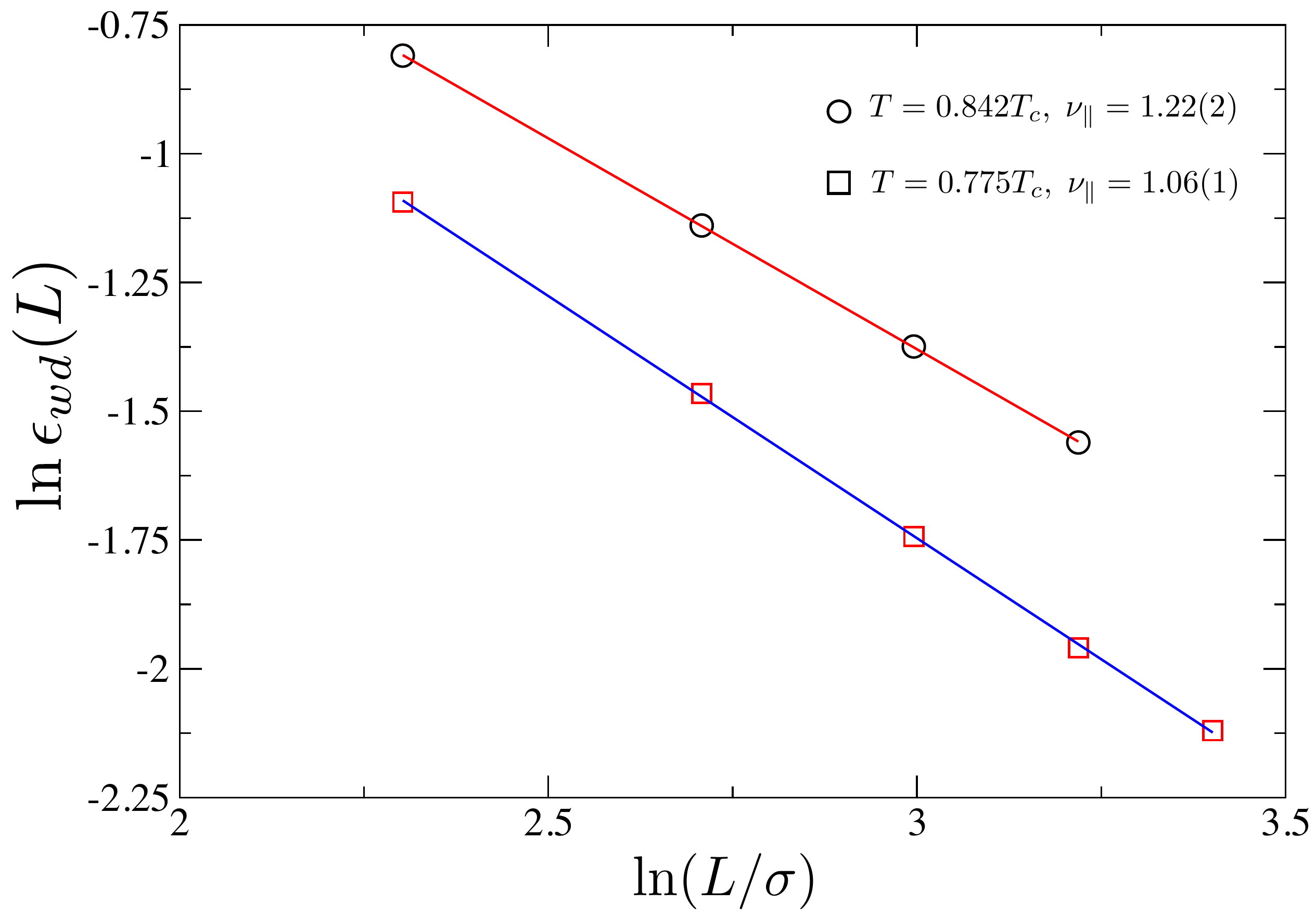}
\caption{The scaling of $\epsilon_{wd}(L)$, i.e. the wall strength at
    which a peak appears in $P(\rho)$, as a function of $L$ for the LR
    wall-fluid potential at vapor-liquid coexistence. Data are shown for two subcritical
    temperatures.}
\label{fig:lpeaks}
\end{figure}

We have determined the value of $\nu_\parallel$ via the anticipated FSS $\epsilon_{wd}(L)\sim
L^{-1/\nu_\parallel}$; $\epsilon_{wd}=0$ for the LR case. For a number of choices of $L$ we measured
$\epsilon_{wd}(L)$ accurately (via histogram extrapolation) from the vanishing of the liquid peak of $P(\rho)$ (cf.
Fig.~\ref{fig:LRFS}). As Fig.~\ref{fig:lpeaks} shows, we do indeed see power law scaling, from which we can
extract an estimate of $\nu_\parallel$  --see key. Interestingly, however, this estimate exceeds the prediction
$\nu_\parallel=0.5$ of mean field and RG theories (see Secs.~\ref{sec:bp}, \ref{sec:rg}) by more than a factor of
two and additionally appears to show a clear dependence on the temperature.

\begin{figure}[h]
\includegraphics[type=pdf,ext=.pdf,read=.pdf,width=0.94\columnwidth,clip=true]{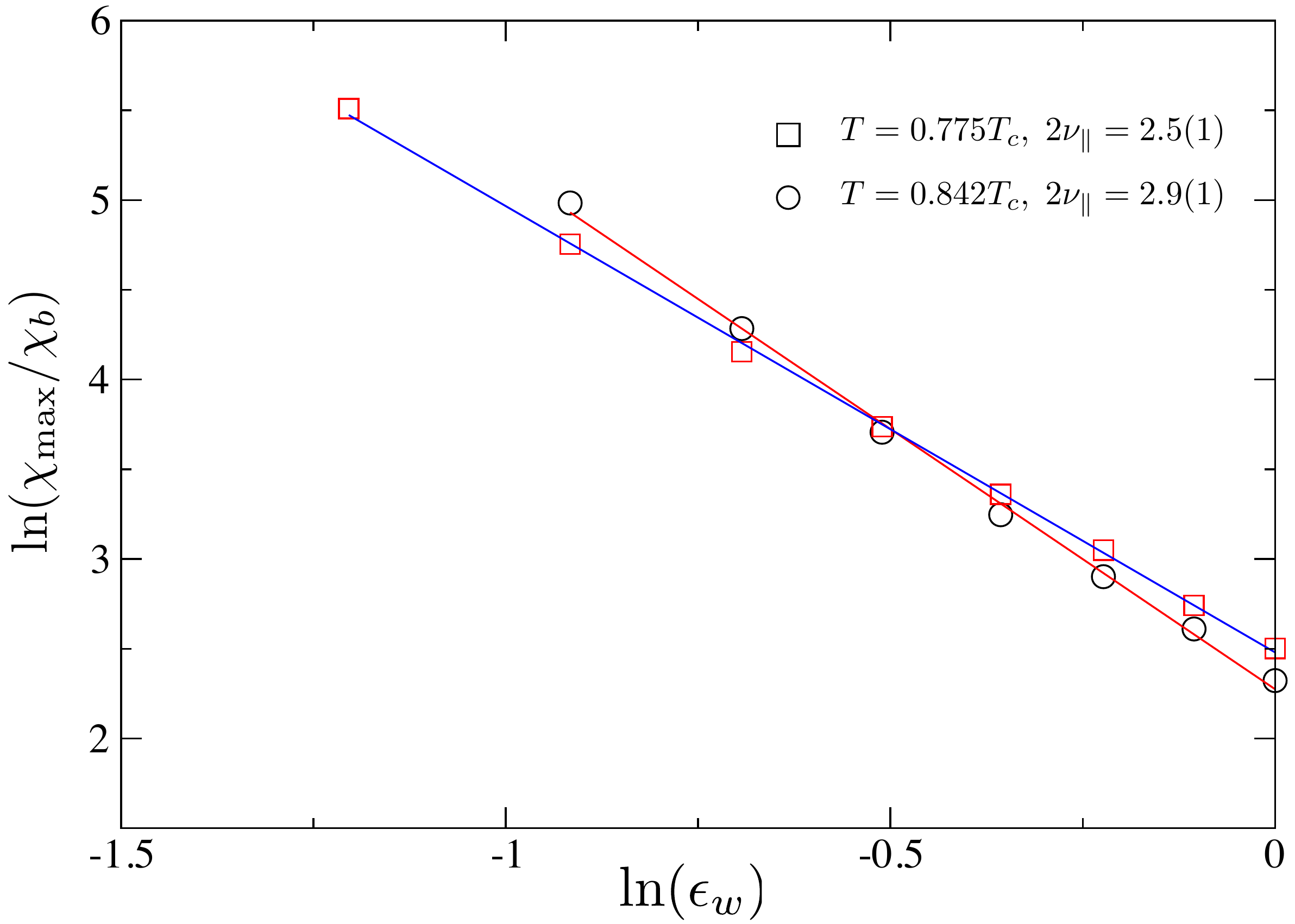}
\caption{GCMC measurements of the scaling of the peak in
    $\chi(z)/\chi_b$ with wall strength $\epsilon_w$ for
    the LR wall-fluid potential at vapor-liquid coexistence. $\chi_b$ is the bulk liquid phase
    compressibility. The system size is $L=50\sigma, D=30\sigma$ and data are shown for two subcritical
    temperatures.}

\label{fig:comp_peak_height}
\end{figure}

A  further independent estimate of $\nu_\parallel$ can be obtained from the growth in the maximum of $\chi(z)$ on
the approach to critical drying. On theoretical grounds one expects (cf. Eq.~(\ref{eq:chizS})) that $\chi_{\rm
max}\sim(\epsilon_w-\epsilon_{wd})^{-2\nu_\parallel}$. The data of Fig.~\ref{fig:comp_peak_height}, when fitted to a power law, yield an
estimate of $\nu_\parallel$ (see key of Fig.~\ref{fig:comp_peak_height}) that is again more than twice the theoretical
prediction and (in common with the finding for $\epsilon_{wd}(L)$) seem to demonstrate a clear temperature
dependence. We shall return to these findings in Sec.~\ref{sec:discuss}.

\subsection{Critical wetting}

\begin{figure}[h]

\includegraphics[type=pdf,ext=.pdf,read=.pdf,width=0.94\columnwidth,clip=true]{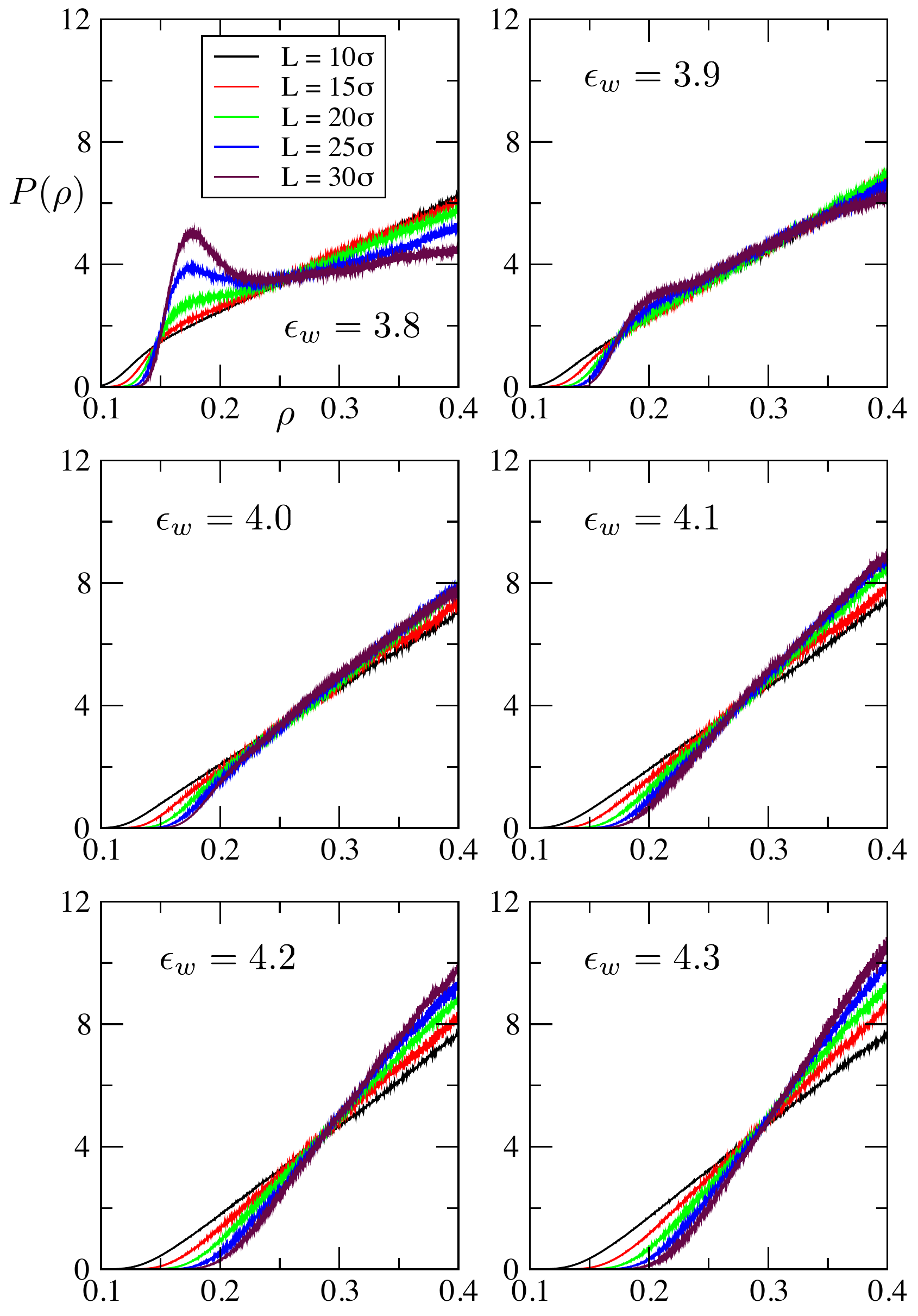}
\caption{GCMC results for $P(\rho)$ for the SR wall
    potential for $D=30\sigma$ and various $L$ at a selection of
    wall strengths $\epsilon_w$ near to the critical wetting point.}
\label{fig:critwet}
\end{figure}

So far we have focused exclusively on critical drying. However the contact angle measurements of
Figs.~\ref{fig:costhetadft} and \ref{fig:costhetasim} suggest that for a SR wall-fluid potential (\ref{eq:SRpot}), critical
wetting occurs. It is therefore of interest to ask how the associated phenomenology compares with that of critical
drying. To this end we have performed simulation studies of the $L$-dependence of $P(\rho)$ in the
neighborhood of the wetting point suggested by the contact angle measurements. Our results are
shown in Fig.~\ref{fig:critwet} and exhibit closely analogous behaviour to that seen for critical drying in the SR 
case (Fig.~\ref{fig:SWFS}). The main difference is that here it is the vapor phase that is metastable, displaying a
peak that decays smoothly into a plateau before turning into a bulge, which decays further with increasing
$\epsilon_w$ until all that remains is a linear part and a tail. As for drying, the effect of
increasing $L$ is to increase the height of any peak or the strength of any bulge. However once critical
wetting is reached, $P(\rho)$ exhibits a linear part and a tail for all $L$. The smallest value of $\epsilon_w$
for which this occurs therefore marks the critical wetting point. For our SR system this appears to occur at
$\epsilon_{ww}\approx 4.2(2)$. We note that this is very different from the estimate of $\epsilon_{ww}(L)=3.7
(1)$ which emerges from the contact angle measurements for $L=15\sigma$ and which corresponds to the point at which 
the vapor peak disappears into the plateau. Clearly, therefore, an analysis of finite-size effects is indispensable 
when seeking to obtain accurate estimates of critical wetting points, as indeed it is for critical drying.

\begin{figure}[h]
\includegraphics[type=pdf,ext=.pdf,read=.pdf,width=0.94\columnwidth,clip=true]{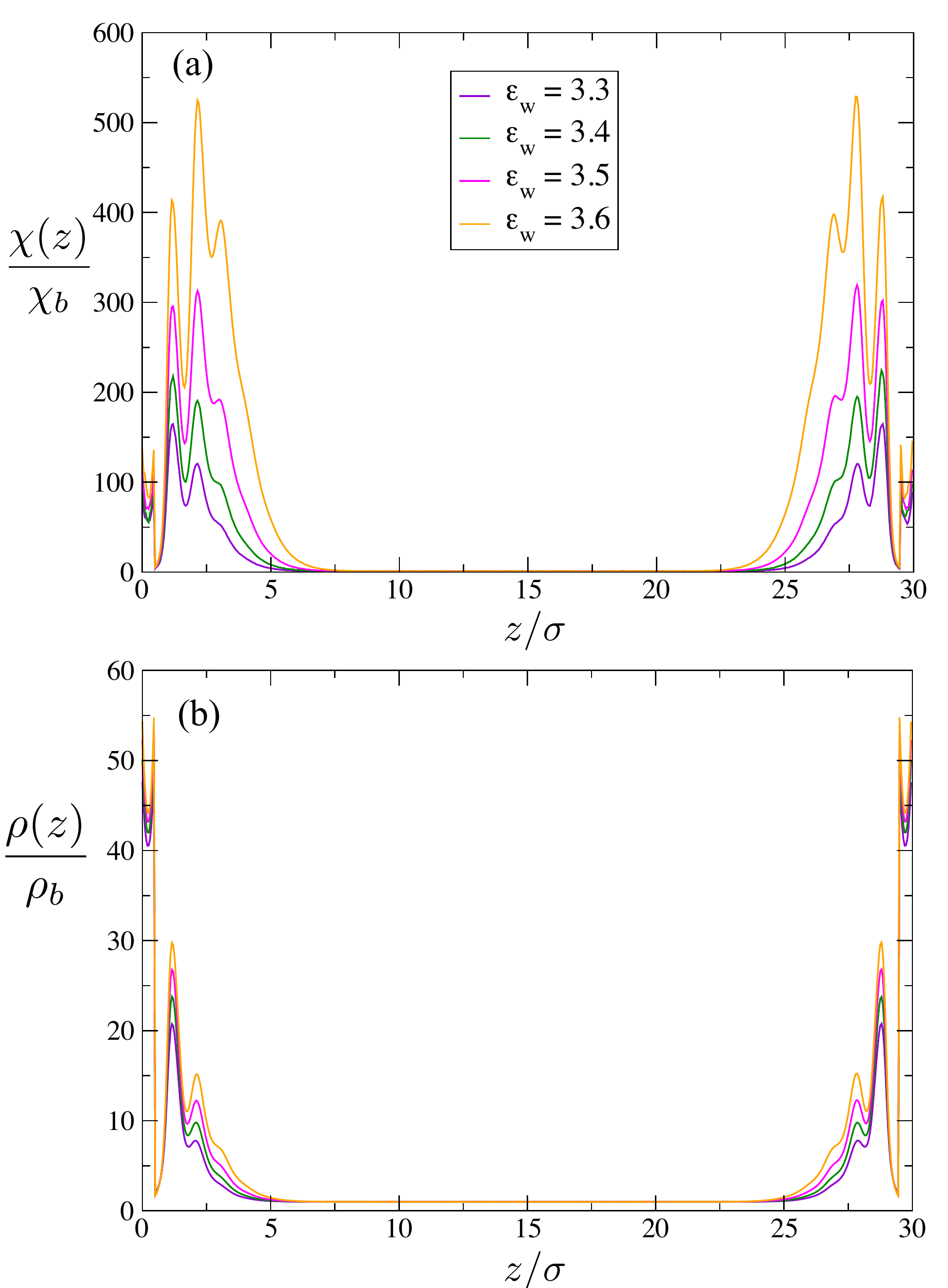}
\caption{{\bf (a)} GCMC results for the normalised compressibility profile $\chi(z)/\chi_b$  at vapor liquid coexistence for various strengths $\epsilon_w$ of the SR system, approaching critical wetting. The system size is $L=50\sigma,D=30\sigma$, and the temperature is $=0.775T_c$.  {\bf (b)}  Corresponding results for the normalised density profile $\rho(z)/\rho_b$.}
\label{fig:wetprofs}
\end{figure}

In Fig.~\ref{fig:wetprofs} we show results for the normalised compressibility profile $\chi(z)/\chi_b$ and density
profile $\rho(z)/\rho_b$ on the approach to critical wetting. For the system size $L=50\sigma$ that we used, the
estimates were unaffected by finite-size effects for $\epsilon_{w}\lesssim 3.6$. In common with the results for critical
drying (cf. Fig.~\ref{fig:profiles}), one sees a very large relative compressibility near the walls. The local
compressibility in the liquid-like layers near the walls is up to 500 times that of the bulk vapor at the same
coexistence chemical potential. The main difference between critical drying and wetting is that owing to the very strong
SR wall attraction in the case of wetting, packing effects occur in the density profile near the wall, and these modulate
the compressibility profile as well. An interesting feature of the density profile, Fig.~\ref{fig:wetprofs}(b), is that
a very dense single layer of particles is strongly adsorbed on the walls. Given the very short range ($0.5\sigma$) of
the square well $wf$ potential, particles beyond this layer do not interact with the wall directly, rather they
experience an effective confining potential that stems from the dense first layer, and whose form is that of the
truncated LJ $ff$ potential. This implies the system is still governed by SR interactions, allowing critical wetting to
occur.

\section{Discussion and conclusions}
\label{sec:discuss}

In summary, we have investigated the properties of a fluid having truncated particle-particle ($ff$) interactions which
is confined between smooth planar walls. Two forms of the wall-fluid interaction potential $W(z)$ have been considered:
the long ranged (LR) case (\ref{eq:LRpot}) in which $W(z)$ exhibits power law decay and the short ranged (SR) case
(\ref{eq:SRpot}) of a square well potential. Clear evidence has been obtained that the character of the wetting and
drying transitions is sensitive to this range. In particular we find from simulation and DFT that for the LR case
wetting is first order, while for the square well wall it is critical \footnote{In a further investigation whose results
we do not show here, we have studied the effect of truncating the $9$-$3$ {\em wf} potentials at different cut-off
distances $z_{cw}$ in (\ref{eq:LRpot}). As expected, we find that the wetting transition is strongly first order for
large values of the cut-off. It remains first order as $z_{cw}$ is decreased to about $2\sigma$. For smaller values the
transition appears to be critical, as in the square well case.}. By contrast, drying is critical for both the SR and LR
cases. For the latter case drying occurs at exactly zero attractive wall strength ie. for a hard wall. Of course it is
well known that in this limit complete drying occurs for all $T<T_c$ \cite{Henderson:1985aa,Oettel:2005aa}. What is
remarkable, however, is that the transition is critical and is predicted to occur precisely at $\epsilon_w=0$ for all LR
(power law) forms of $W(z)$.

Knowledge of the exact location of a surface phase transition in 3d provides a unique opportunity to study a surface
critical point free from uncertainty regarding its location (a problem that has previously plagued Ising model studies
of critical wetting \cite{Bryk:2013pi}). We have obtained simulation estimates of the surface critical exponent
$\nu_\parallel$ in the LR case and compared these with the predictions of mean field theory and a linear renormalization
group calculation, finding our estimate to be over twice the predicted value. Furthermore, the exponent estimates exhibit a clear temperature
dependence -- a feature which is also at variance with the theoretical predictions. Given that $d=3$ is the upper critical dimension
for this system, at which mean field theory for the exponents is expected to hold, these findings are unexpected.

A possible reason for the discrepancy is finite-size effects, the nature of which we have sought to elucidate.  
Finite-size effects for surface criticality have been investigated previously in the context of critical wetting in the Ising
model with SR forces (where the special symmetry implies that wetting and drying are
equivalent). A finite-size scaling ansatz proposed by Binder and co-workers \cite{Albano:2012db,Bryk:2013pi}
assumes that critical wetting in 3d can be treated on the same footing as a bulk transition in which
two independent correlation lengths diverge in orthogonal directions. However, our results call this assumption
into question. We find that while $\xi_\parallel$ can grow as large as the wall
dimension $L$, $\xi_\perp$ is, by contrast, heavily damped for finite $L$ and remains microscopic for
all accessible system sizes (recall that capillary wave theory predicts $\xi_\perp\sim\sqrt{\ln L}$ in $d=3$). Moreover,  as $\xi_\parallel\to L$ the system spontaneously
exits the near-critical state due to the premature unbinding of the metastable phase. This state of
affairs differs qualitatively from that for bulk criticality where critical
fluctuations can be measured right up to the critical point, together with quantities such as cumulants of the order
parameter distribution. Clearly fresh and bespoke FSS approaches are need for dealing with surface criticality in fluid systems.

Given the heavy dampening of $\xi_\perp$, the surface critical behaviour observed in simulations appears to be
controlled by the single diverging lengthscale $\xi_\parallel$. It is therefore tempting to speculate that effective
critical surface behaviour is $2d$ Ising like in character. Indeed our measured values of $\nu_\parallel$ are much
closer to $\nu_\parallel=1$ than they are to the theoretical prediction $\nu_\parallel=1/2$.  Another issue to note is that owing to its
slow (logarithmic) divergence as $\epsilon_w\to0$, the thickness of the drying layer does not exceed a few particle
diameters at the state point closest to criticality for which we can attain the thermodynamic limit. Of course, for very
large system sizes one might expect to observe a crossover to the mean-field and RG predictions of
Sec.~\ref{sec:theory}; however no hints of such a crossover are visible in our results, despite the investment of
substantial computational resources to study large systems. In our view, if such a crossover exists, there seems little
hope of observing it in the foreseeable future.

Our finding that premature drying occurs when $\xi_\parallel\simeq L$ helps to explain a longstanding
controversy in the literature
\cite{Swol:1989by,Henderson:1990nq,Swol:1991fq,Nijmeijer:1992fk,Nijmeijer:1991sw,Henderson:1992kk,Bruin:1995ud,Bruin:1998aa}
concerning the order of the drying transition. On the basis of MD simulation, van Swol and Henderson asserted
that the drying transition for a square well fluid with SR wall-fluid interactions is first order in
character. The evidence for this was that at small attractive wall strength an abrupt change was observed in the density
profile $\rho(z)$ from a liquid-like profile to a gas-like one. Given the insights provided by the present
work, one can see that rather than being associated with a first order drying transition, this phenomenology
arises from the premature unbinding of the liquid slab which then diffuses away from wall. 

One of the motivations for the present work was to contribute to ongoing attempts to understand the correlation between
the structure of water near a hydrophobic surface and the value of the contact angle. Experimental studies have reported
a region of depleted density in the close proximity of hydrophobic surfaces \cite{Mezger:2006zl, Ocko:2008fv,
Mezger:2010lq,Chattopadhyay:2010aa,Chattopadhyay:2011aa,Uysal:2013aa}, while simulation studies report a growth in
density fluctuations near the surface as the contact angle increases \cite{Chandler:2007aa,
Patel:2010dz,Patel:2012aa,Mittal:2010aa,Willard:2014aa,Godawat:2011aa}. As we have shown previously \cite{Evans:2015wo},
both phenomena in water can be accounted for if hydrophobic substrates can be associated with the approach to the
critical drying point. Pertinent is the form of the wall-oxygen potential. In the GCMC simulations of
Ref.~\cite{Kumar:2013aa} this was non-truncated $9$-$3$, equivalent to (\ref{eq:LRpot}), and the water model was SPC/E
which is SR. On the basis of the arguments presented here we would predict critical drying of water in the limit where
the attraction strength $\epsilon_w$ vanishes. Indeed Kumar and Errington (see Fig. 9 of \cite{Kumar:2013aa}) appeared
to observe $\cos(\theta)$ approaching $-1$ (tangentially) for very small $\epsilon_w$. In our own study of SPC/E
\cite{Evans:2015wo} the $9$-$3$ wall-oxygen potential was truncated at a large distance $15$\AA ~and similar behaviour
was found but we could not identify the drying point accurately. The MD study \cite{Willard:2014aa} employed a
wall-oxygen potential with a non-truncated $z^{-6}$ tail. Once again we would predict critical drying in the limit of
vanishing attraction. It is likely that the strong density fluctuations close to the wall that are observed in
\cite{Willard:2014aa} are associated with the approach to the critical region; the authors consider very weak $wf$
attraction but their estimates of $\cos(\theta)$ are not sufficiently accurate to address the location of the transition
point. The present study of a generic fluid model serves to emphasize that there is nothing special about water with
respect to these phenomena. Indeed it seems that critical drying should be expected for all liquids provided the
substrate is sufficiently weak and this will be associated with a growing drying layer and enhanced density fluctuations
near the substrate. Furthermore it can be expected that the presence of the critical drying point is to be felt
throughout the hydrophobic (partial drying) regime i.e. not just in the limit $\cos(\theta)=-1$.
  
Finally we briefly point out a number of avenues for future work. While the present study has focused on critical drying in the particular case of SR $ff$ with LR $wf$
because here the critical point is known exactly, it would
be interesting to see if one can determine critical properties for drying and wetting in a SR system such as a square well
wall (\ref{eq:SRpot}). Our results (Figs.~\ref{fig:SWFS},\ref{fig:critwet}) strongly suggest that the hallmark of
surface criticality is the same as for the LR case, namely a scale invariant limit in which $P(\rho)$ exhibits a linear
part with a tail. It remains to be seen whether application of this method for locating criticality provides sufficient
accuracy to allow reliable estimates of exponents, which are predicted to depend on the form of the $wf$ and
$ff$ potentials-see Sec.~\ref{sec:phenomenology}. Beyond this, it would also be interesting to investigate the case most pertinent to
real systems, where dispersion forces ensure that both $wf$ and $ff$ are LR and critical drying can occur at non-zero
wall strength (whose value is known exactly in terms of the coexisting densities
\cite{Stewart:2005qd}) and the critical exponents are predicted to take large values, e.g. $\nu_\parallel=5/2$
\cite{Dietrich:1988et}. 

\acknowledgments
We have benefited from helpful discussions with A.J.~Archer and A.O.~Parry.
R.E. acknowledges Leverhulme Trust grant EM-2016-031. This research made 
use of the Balena High Performance Computing Service at the University of Bath.

\bibliography{/Users/pysnbw/Dropbox/Papers}

\end{document}